\documentclass[namedreferences,aas_macros]{solarphysics}

\usepackage[hyperref,optionalrh,solaromanenum]{spr-sola-addons} 
\usepackage{graphicx}                    
\usepackage{xcolor}                       

\usepackage{epstopdf}
\usepackage{lscape}
\usepackage{pdflscape}
\usepackage{wrapfig}
\usepackage{lineno}
\usepackage{changepage}
\usepackage[normalem]{ulem}

\newcommand{\degree}{$^{\circ}$~}
\newcommand{\rsun}{~R_{\odot}}

\newcommand\gtx[1]{{\color{green}#1}}

\begin{document}

\begin{article}

\begin{opening}


\gtx{\title{Comprehensive Characterization of Solar Eruptions With Remote and {\it In-Situ} Observations, and Modeling: The Major Solar Events on 4 November 2015}}

\author[addressref={aff1},email={iver.cairns@sydney.edu.au,+61 2 9351 3961},corref]{Iver H.~Cairns$^{\star}$}
\author[addressref={affkamen},email={kkozarev@astro.bas.bg}]{Kamen~A.~Kozarev}
\author[addressref={affnariaki},email={nitta@lmsal.com}]{Nariaki~V.~Nitta}
\author[addressref={affneus},email={agueda@fqa.ub.edu}]{Neus~Agueda}
\author[addressref={affuclan,affutu},email={markus.battarbee@gmail.com}]{Markus~Battarbee}
\author[addressref={affeoin},email={eoincarley@gmail.com}]{Eoin P.~Carley}
\author[addressref={affnina},email={dresing@physik.uni-kiel.de}]{Nina~Dresing}
\author[addressref={aff99},email={raul.gomezh@uah.es}]{Ra\'ul~G\'omez-Herrero}
\author[addressref={aff2},email={ludwig.klein@obspm.fr}]{Karl-Ludwig~Klein}
\author[addressref={afflario2,afflario},email={david.larioloyo@nasa.gov}]{David~Lario}
\author[addressref={affjens},email={jens.pomoell@helsinki.fi}]{Jens~Pomoell}
\author[addressref={aff2,affcarolina},email={carolina.salas@cinespa.ucr.ac.cr}]{Carolina~Salas-Matamoros}
\author[addressref={aff10},email={astrid.veronig@uni-graz.at}]{Astrid~M.~Veronig}
\author[addressref={aff1},email={bo.li@sydney.edu.au}]{Bo~Li}
\author[addressref={aff1},email={mccauley.pi@gmail.com}]{Patrick~McCauley}
%
\runningauthor{ISSI team: I.H.~Cairns {\it et al.}}
\runningtitle{Events of 4 November 2015}

\address[id={aff1}]{School of Physics, University of Sydney, NSW 2006, Australia}
\address[id={affkamen}]{Institute of Astronomy, Bulgarian Academy of Sciences, 1478 Sofia, Bulgaria}
\address[id={affnariaki}]{Lockheed Martin Advanced Technology Center, Palo Alto, CA 94304 USA}
\address[id={affneus}]{Dep. F\'isica Qu\`antica i Astrof\'isica, Institut de Ci\`encies del Cosmos (ICCUB), Universitat de Barcelona, 08028 Barcelona, Spain}
\address[id={aff2}]{LESIA, Observatoire de Paris, PSL Research University, CNRS, Sorbonne Universit\'{e}s, UPMC Univ. Paris 06, Univ. Paris Diderot, Sorbonne Paris Cit\'{e}, 5 place Jules Janssen, 92195 Meudon, France}
\address[id={affnina}]{Institut f\"uer Experimentelle und Angewandte Physik, University of Kiel, Germany}
\address[id={aff99}]{Space Research Group, Dpto. de F\'isica y Matem\'aticas, University of Alcal\'a, 28871 Alcal\'a de Henares, Madrid, Spain}
\address[id={affeoin}]{Astrophysics Research Group, School of Physics, Trinity College Dublin, Dublin 2, Ireland}
\address[id={afflario}]{NASA Goddard Space Flight Center, Heliophysics Science Division, Greenbelt, MD 20771, USA}
\address[id={afflario2}]{formerly at Applied Physics Laboratory, The Johns Hopkins University, Laurel, MD 20723, USA}
\address[id={affjens}]{Department of Physics, University of Helsinki, Helsinki, Finland}
\address[id={affuclan}]{Jeremiah Horrocks Institute, University of Central Lancashire, PR1 2HE, UK}
\address[id={affutu}]{University of Turku, Turku, Finland}
\address[id={affcarolina}]{Space Research Center, University of Costa Rica, San Jos\'e, Costa Rica}
\address[id={aff10}]{Kanzelh\"ohe Observatory \& Institute of Physics, University of Graz, 8010 Graz, Austria}

\begin{abstract}
Solar energetic particles (SEPs) are an important product of solar activity. They are connected to solar active regions and flares, coronal mass ejections (CMEs), EUV waves, shocks, Type II and III radio emissions, and X-ray bursts. These phenomena are major probes of the partition of energy in solar eruptions, as well as for the organization, dynamics, and relaxation of coronal and interplanetary magnetic fields. Many of these phenomena cause terrestrial space weather, posing multiple hazards for humans and their technology from space to the ground. Since particular flares, shocks, CMEs, and EUV waves produce SEP events but others do not, since propagation effects from the low corona to $1$~AU appear important for some events but not others, and since Type II and III radio emissions and X-ray bursts are sometimes produced by energetic particles leaving these acceleration sites, it is necessary to study the whole system with  a multi-frequency and multi-instrument perspective that combines both {\it in-situ} and remote observations with detailed modelling of phenomena. This article demonstrates this comprehensive approach, and shows its necessity, by analysing a trio of unusual and striking solar eruptions, radio and X-ray bursts, and SEP events that occurred on 4 November 2015. These events show both strong similarities and differences from standard events and each other, despite having very similar interplanetary conditions and only two flare sites and CME genesis regions. They are therefore major targets for further in-depth observational studies, and for testing both existing and new theories and models. We present the complete suite of relevant observations, complement them with initial modelling results for the SEPs and interplanetary magnetic connectivity, and develop a plausible scenario for the eruptions. Perhaps controversially, the SEPs appear to be reasonably modelled and evidence points to significant non-Parker magnetic fields. Based on the very limited modelling available we identify the aspects that are and are not understood, and we discuss ideas that may lead to improved understanding of the SEP, radio, and space-weather events.
\end{abstract}

%
\keywords{solar energetic particles (SEPs), flares, coronal mass ejections, EUV waves, radio bursts, X-ray bursts, magnetic field, particle propagation}
\end{opening}

%

\section{Introduction}
\label{S1}

Solar flares and coronal mass ejections (CMEs) result from magnetic reconnection changing magnetic topologies and releasing energy from magnetic loops and active regions produced by the Sun's magnetic dynamo. Sufficiently energetic flares and CMEs can produce large-scale propagating waves, most plausibly in the (magnetosonic) fast-mode and Alfv\'en modes, and pulse-like disturbances. Examples include ``EUV waves'' observed in the lower corona at EUV wavelengths \citep{thompson_etal_1998, warmuth_2007, veronig_etal_2010, patsourakos_and_vourlidas_2012, webb_and_howard_2012}, originally called EIT waves \citep{thompson_etal_1998}, and Moreton waves observed in the chromosphere using H$_{\alpha}$ \citep{moreton_1960,uchida_1968}. CMEs and other sufficiently fast plasma motions -  not necessarily faster than the fast-mode speed \citep{pomoell_etal_2008} - can lead to fast-mode waves that steepen nonlinearly into shocks. Usually idealised as abrupt discontinuities, these shocks compress, heat, and alter the bulk velocity of the plasma, amplify and rotate the magnetic field, and can accelerate particles. CME-driven shocks are visible directly in white-light observations \citep{vourlidas_etal_2003,schmidt_etal_2016}. Since shocks convert the kinetic energy of a disturbance into thermal, magnetic, and accelerated particle energy, driven shocks are expected to persist longer than blast-wave shocks, for which the shock has propagated well away from the driver.  
  
Originally defined by near-Earth space observations, solar energetic particles (SEPs) are produced between the Sun and Earth as a result of solar activity, as reviewed for example by \citet{reames_1999} and \citet{klecker_etal_2006}. SEPs have several important space-weather consequences, including radiation damage to technological systems ({\it e.g.} degradation of solar cells and electrical circuit components) and humans ({\it e.g.} astronauts and air travelers), modifying the Earth's radiation belts and environment \citep{baker_etal_2013}, and causing increased particle precipitation into the ionosphere, with associated changes in ionization, plasma density, and radio propagation effects. 

The general importance of SEPs and their many associated solar and interplanetary phenomena is that they involve physics that is fundamental, unusually well observed (with high temporal resolution remote imaging data from gamma rays to radio waves, plus {\it in-situ} particle, magnetic-field, and wave observations), and also widely applicable across the fields of astrophysics, plasma physics, and space physics. For instance, SEP production and propagation involves the acceleration of particles in reconnection regions and by shocks and turbulence, the scattering of particles by magnetic (and electric) field turbulence and self-generated waves, the evolution and dynamics of turbulence, interplanetary magnetic-field connectivity, and the propagation and evolution of CMEs and shocks in the corona and solar wind. Similarly, electrons that are accelerated in solar flares and move down towards the chromosphere lead to (reverse drift) Type III solar radio bursts and X-rays, while the associated downward-going ions produce gamma-rays and X-rays. Precipitation of these particles leads to chromospheric heating, expansion, and evaporation fronts. Electrons accelerated outwards lead to (normal drift) Type III bursts in the corona and solar wind, as well as the prompt component of SEPs; the corresponding ions become prompt SEPs. Shocks, whether blast waves produced by flares or CME-driven shocks, contribute strongly to SEPs; the electrons also produce Type II solar radio bursts in the corona and/or solar wind. Finally, Moreton waves, EUV waves, and EIT waves are signatures of dynamic activity that are sometimes associated with SEP acceleration.

Multiple unresolved issues exist concerning the production and propagation of SEPs from the Sun to Earth, their association with flares, CMEs, and the multiple signatures of activity summarized above, and the physics of these signatures themselves. It is plausible that a definitive answer to the questions of whether and how efficiently coronal shocks accelerate SEPs will require carefully combining in-situ and remote sensing observations with realistic global modelling ({\it e.g.} \citet{lario_etal_2017a}. These observations and models will not only be for SEPs but also for the related phenomena of flares, erupting filaments and CMEs, EUV and Moreton waves, and Type II and III bursts. Some of these, especially the propagation and properties of CMEs and their white-light/EUV and Type II radio signatures, are also relevant to forecasting space weather \citep{schmidt_etal_2013, cairns_and_schmidt_2015, kozarev_etal_2015, schmidt_and_cairns_2015, schmidt_and_cairns_2016}. Resolving these issues is a major focus of the {\it Parker Solar Probe} \citep{fox_etal_2016} and {\it Solar Orbiter} \citep{muller_etal_2013} missions.

In this article we briefly review and summarise these issues and then address them using the major solar and interplanetary events associated with the events of 4 November 2015. Arguably these are an ideal set of events to study. First, all three are major events that occur in essentially the same coronal and interplanetary configurations ({\it e.g.} the same large-scale magnetic connectivity and structures such as streamers and coronal holes) and conditions (except for seed energetic particles) but with two sites for the flares and associated CMEs. Both these are on the side of the Sun facing Earth, one near the west limb and one near disk center. Second, a very complete set of ground-based and spacecraft observations exists, ranging from remote X-ray to radio wavelengths for light to {\it in-situ} plasma, field, and energetic particle measurements, making these events amenable to comprehensive data-theory comparisons. Third, significant SEP, space-weather, flare, CME, EUV wave, Moreton wave, hard X-ray, microwave, and radio events were produced, all of direct interest for observers, theorists, modelers, simulators, and operational space-weather staff. What is more, a relatively complete set of observations exists for these. Fourth, the events show strong commonalities (M-class flares, observable EUV waves, CMEs, and Type II and III bursts), yet also strong differences (magnetic connectivity, SEP occurrence, and radio bursts. Fifth, an unusually strong space-weather event occurred in association with the third event on 4 November 2015, especially with regard to aviation radar systems \citep{marque_etal_2018}.

Our primary goal is to detail the solar and interplanetary observations for the three events, describing the common and different features, identifying the aspects that are and are not understood now, and providing the basic observations in a form amenable for future, more detailed, comparisons with theoretical and modeling analyses. The secondary goal is to make progress on understanding these solar and interplanetary phenomena, especially those associated with SEPs, shocks, and magnetic-field configurations, by showcasing the necessary elements of a comprehensive analysis including multi-instrument observations and relevant modeling.

The article proceeds by reviewing the issues involved with SEPs, flares, shocks, CMEs, waves, and related signatures (Section \ref{S2}). It then describes the evolution of the parent active regions, coronal magnetic field, and the X-ray and microwave flares for the 4 November 2015 events (Section \ref{S3}). The failed filament eruptions for the first two flares, the Moreton wave for the first flare, and the EUV waves and CMEs for all three flares are described in Section \ref{S4} and shown to be mutually consistent. Section \ref{S5} overviews the radio events, including the properties of the over five Type IIs involved, the relative lack of Type IIIs, and the strong microwave and Type IV emission for the first and third events. Section \ref{S6} details the interplanetary plasma and magnetic field context, showing the arrival of a shock early on 4 November associated with an earlier event and the shock and CME associated with the third event. The SEP observations are detailed in Section \ref{S7}. The space-weather aspects of the events are briefly discussed in Section \ref{S8}. A summary of the observations and associated theoretical implications is provided in Section \ref{S9}.

\section{Detailed Theoretical and Observational Context}
\label{S2}

SEPs consist of electron, proton, and ion populations with energies in the range of tens of KeV to a few GeV. SEP events can be loosely categorized into impulsive and gradual events, distinguished by the timescales of their intensity profiles and properties such as composition and ionization states \citep{ luhn_etal_1984,cane_etal_1986,reames_stone_1986,reames_1988,luhn_etal_1987,reames_1999, klecker_etal_2006}. Impulsive events are attributed to particle acceleration in regions producing solar flares, presumably in magnetic-reconnection regions, and gradual events to acceleration by CME-driven shocks. However, a number of events exhibit characteristics of both impulsive and gradual events ({\it i.e.} timing of intensity profiles and ratios of heavy ions at high energies with hybrid characteristics), blurring the distinction between acceleration at reconnection sites and at shocks \citep{kallenrode_etal_1992, torsti_etal_2002, Klecker_Mobius_Popecki_2006}.

{\it In-situ} observations of CME-driven shocks and their associated energetic particles have shown that particle acceleration at shocks typically results from shock drift acceleration in the quasi-perpendicular regime and diffusive shock acceleration in the quasi-parallel regime \citep[{\em e.g.}][and references therein]{lee_1983,decker_vlahos_1986,kennel_etal_1986,jones_ellison_1991,lee_2005,cohen_2006,desai_giacalone_2016}. Here the two regimes are defined in terms of $\theta_{Bn}$, the angle between the upstream magnetic field ${\bf B}$ and the normal to the local shock surface: the quasi-perpendicular and quasi-parallel regimes correspond to $45^{\circ} \lesssim \theta_{Bn} \lesssim 90^{\circ}$ and $\theta_{Bn} \lesssim 45 ^{\circ}$, respectively. Without {\it in-situ} measurements of shocks and magnetic fields in the corona, determining which acceleration processes take place close to the Sun is very challenging and requires careful examination of multi-wavelength remote observations.

Analyses of white light and EUV observations, supported by radio imaging and radio Type II dynamic spectra, have found that shocks can form as low in the corona as heliocentric distances of $1.2$ to $2.2 \rsun$   \citep{klassen_etal_1999,veronig_etal_2010,Ma2011,bain_etal_2012,gopalswamy_etal_2013,carley_etal_2013,nitta_etal_2014}, where $\rsun$  denotes the solar radius. Several mechanisms can give rise to shock formation in the corona, including blast waves caused by a sudden release of flare-related energy \citep{Vrsnak_etal_2006} and erupting CMEs which drive shocks as they propagate outwards \citep{dauphin_etal_2006, zimovets_etal_2012}. Determining whether blast-wave or CME-driven shocks are relevant to particular events \citep{howard_pizzo_2016}, and especially to associated Type II bursts and EUV waves, is of particular interest \citep{cane_and_erickson_2005, cairns_2011}. Essentially all interplanetary Type II bursts are interpreted in terms of CME-driven shocks \citep{reiner_etal_1998, bale_etal_1999, cairns_2011}, but this may not be correct for coronal Type IIs. 

Type II bursts are interpreted theoretically in terms of: shock-drift acceleration and magnetic mirror reflection of electrons at shocks; development of a beam distribution of reflected electrons; growth of Langmuir waves via the beam instability; and nonlinear wave--wave processes that convert Langmuir wave energy into radio emission near the electron plasma frequency $f_{pe}$ and near $2f_{pe}$ (the so-called fundamental and harmonic radiation, respectively). Relevant reviews include those of \cite{nelson_and_melrose_1985}, \cite{bastian_etal_1998}, and \cite{cairns_2011}. These theories require the source regions to be where the shock is strongly quasi--perpendicular with 80\degree $\lesssim \theta_{Bn} \lesssim$~90\degree \citep{holman_and_pesses_1983, cairns_1986, knock_etal_2001, schmidt_and_cairns_2012b, cairns_and_schmidt_2015, schmidt_and_cairns_2016}. Interestingly, multi-frequency mapping of some Type II bursts shows that source regions at different frequencies can be aligned along a direction that is strongly inclined to the radial \citep{Nelson_and_Robinson_1975, Klein_etal_1999}. This is not expected if the electrons are produced at quasi-perpendicular regions of the shock ({\it e.g.} near the nose for overlying loop fields or at lateral expanding flank regions for quasi-radial fields) or at quasi-parallel regions of the shock ({\it e.g.} near the nose for quasi-radial ${\bf B}$). Recent semi-empirical studies \citep{kozarev_etal_2015, lario_etal_2017a} have suggested that the regions of expected shock acceleration may vary with time, and may move to different locations on the shock surface, depending on the parameters governing acceleration efficiency. Combining remote observations with modeling approaches allows determination of relevant parameters for electron and ion acceleration: $\theta_{Bn}$, the spatial profile of the Alfv\'en speed $V_{A}$, and the lateral expansion of the driving CME \citep{warmuth_and_mann_2005,Temmer_etal_2013, zucca_etal_2014,kozarev_etal_2015,lario_etal_2017a,lario_etal_2017b}.

Recent high-cadence observations of large-scale coronal transients, known as ``EUV waves'' (or ``EIT waves'', ``coronal bright fronts (CBFs)'', and ``large-scale coronal propagating fronts (LCPFs)''), suggest that they are signatures of magnetosonic waves or shocks \citep{warmuth_etal_2004, veronig_etal_2010, kozarev_etal_2011, downs_etal_2012}. Here we consistently use the term ``EUV waves'' to avoid unnecessary confusion. EUV waves have been widely studied in the last several years due largely to the significantly improved EUV images in terms of spatial and temporal resolution, spectral coverage, and multipoint views available from the {\it SOlar and Heliospheric Observatory} (SOHO), {\it Solar and TErrestrial RElations Observatory} (STEREO), and {\it Solar Dynamics Observatory} (SDO). We now know that EUV waves are very common during sufficiently impulsive solar eruptions and several studies have characterized them in detail \citep{veronig_etal_2010, patsourakos_etal_2010, hoilijoki_etal_2013}. 

The ubiquity of EUV waves during solar eruptions has raised the question of whether they signify shocks or compression waves responsible for accelerating particles observed during the early stages of SEP events. Extending classic works \citep{krucker_etal_1999,torsti_etal_1999}, recent analysis of the temporal relation between the evolution of EUV waves on the solar disk and the {\it in-situ} onset of SEP fluxes for a large sample of events during Cycle 23 has shown a general consistency with wave/shock acceleration for protons but not for electrons \citep{miteva_et_al_2014}. Correspondingly, some analyses find evidence for SEP injections when EUV waves reach the magnetic footpoint of the spacecraft \citep{rouillard_etal_2012} whereas others do not   \citep{miteva_etal_2014,lario_etal_2014}. This discrepancy points to the likely complexity of the interactions between the EUV wave, shock (whether blast-wave or CME-driven), CME, flare, and the global coronal magnetic field. 

The two STEREO spacecraft and the near-Earth spacecraft {\it Advanced Composition Explorer} (ACE), SOHO, and {\it Wind} allow study of SEP events from multiple vantage points. Observing the same event from a broad range of longitudes \citep[{\it e.g.}]{dresing_etal_2012,lario_etal_2013,dresing_etal_2014,lario_etal_2014,gomez_herrero_etal_2015} allows us to constrain the longitudinal extent of particle acceleration by shocks and associated magnetic connectivity. For some events very different SEP fluxes and profiles are observed at closely separated spacecraft  \citep{KlassenEtAl16}, for others the entire SEP event is very localised in longitude, and for still others the SEP event is observed at all longitudes. These observations and complementary modeling efforts are beginning to unravel the complexity in time, longitude, energy, and species of particle acceleration and transport through the inhomogeneous coronal and solar wind \citep{PachecoEtAl17, afanasiev_and_vainio_2013, kozarev_etal_2013}, as well as the associations with radio emissions \citep{cane_etal_2002, schmidt_etal_2014a, cairns_and_schmidt_2015, schmidt_and_cairns_2016}.

The properties of the seed-particle distributions incident on the shock (whether from the ambient background, a flare site, or pre-processed by another event) also affect both the shock-accelerated particle distribution functions ({\it e.g.} the ``injected'' particles subject to propagation analyses \citep{ battarbee_etal_2013,agueda_etal_2014}) and related phenomena such as Type II bursts \citep{cairns_etal_2003, knock_etal_2003a, kozarev_etal_2015, schmidt_etal_2016, lario_etal_2017b}. The properties of pre-existing and self-generated turbulence also affect the effectiveness of diffusive shock acceleration  ({\it e.g.} \citet{vainio_etal_2014}).  Similarly, in-situ observations of relativistic electrons and ions yield injection / release times and propagation distances that constrain the locations and duration of acceleration events and the effectiveness of wave-particle scattering and diffusion between the source and observer, {\it e.g.} \citet{agueda_etal_2014}. 

Magnetic connectivity between the SEP source and the observing location is required, unless sufficient cross-field scattering and diffusion exist, for SEPs to be observed. This requires particles to either be accelerated on field lines connecting to the observer or have access to these open field lines \citep{lario_etal_2017b}. Modeling of solar and interplanetary magnetic structures is then required, for instance using PFSS or other approaches such as MHD simulations  \citep{luhmann_etal_2017}, the Archimedean (hereafter Parker) spiral \citep{parker_1958}, or generalized data-driven models \citep{li_etal_2016}.

We return to diagnostics of particle acceleration in flares, clearly vital if the effects of shocks and flares are to be identified, separated, and constrained in detail. Flare signatures are observed at H$_{\alpha}$, white-light, UV to EUV, X-ray, and gamma-ray wavelengths. Flares involve substantial heating (sometimes to over $10$~MK \citep{Lin1981, Caspi2014}), changed magnetic topologies, and particle acceleration. Magnetic reconnection is thus directly relevant but other processes likely contribute to the heating and particle acceleration \citep{Fletcher2011}. The spatial sizes of flaring regions vary widely, from very compact regions ({\it e.g.} the size of low-lying loops) to the size of entire active regions. Similarly the corresponding time-scales and total energy releases also vary widely, from impulsive to long duration and classes A to X, respectively. 

Thermal emission from the heated plasma is one component of flare radiation ({\it e.g.} soft X-rays and UV and EUV radiation). Radiation is also emitted by or as a result of energetic particles precipitating into the chromosphere from higher-up acceleration regions; examples include H$_{\alpha}$ radiation and EUV radiation, as well as X-rays produced by bremsstrahlung from energetic electrons with either thermal or nonthermal distribution functions. The X-ray spectra and timescales of bursts can distinguish between thermal and non-thermal electron populations \citep{Holman2011}. 

Another crucial signature of electron acceleration, but also of connection to open magnetic field lines, are Type III solar radio bursts \citep{suzuki_and_dulk_1985, bastian_etal_1998, li_etal_2008a, reid_and_ratcliffe_2014}. These involve the accelerated electrons developing a beam distribution function by time-of-flight effects, growth of Langmuir waves by the beam instability, and nonlinear coupling of the Langmuir waves to produce $f_{p}$ and $2f_{p}$ radiation. Type IIIs thus are believed to differ from Type IIs in the source of the accelerated electrons, the beam's detailed formation mechanism, and the beam's characteristic speed (speeds greater than $20$ electron thermal speeds {it versus} $3$). Type IIIs have widely different starting and ending frequencies, intensities, and drift rates and can drift to lower and higher frequencies, corresponding conventionally to electrons moving away from (normal frequency drift) and towards (reverse frequency drift) the Sun, respectively. The specific reasons Type IIIs are important for SEP, flare, and CME physics is that they are signatures of open magnetic fields accessible to accelerated electrons, are interpreted in terms of electron acceleration in magnetic reconnection regions, and can lead to SEP particles.

Velocity dispersion analyses of the energetic electrons in Type III and SEP events yield injection times and estimated propagation distances (presumably along ${\bf B}$ and assuming negligible energy losses) \citep{lin_1985}. The time-varying pitch-angle distributions can also be compared with theoretical predictions and used to constrain the timing, number, and relative sizes of injections of energetic particles and the transport conditions along the observer's magnetic field line(s) \citep{agueda_etal_2014}. These constraints can then be compared with independent arguments based on the timing, spatial locations, and magnetic connectivity of Type II and III bursts, flares, CMEs, and shocks. A major issue with understanding SEP electrons associated with Type IIIs is that the relativistic electrons appear to have injection times that are typically \,10--\,20 minutes later than the sub-relativistic electrons (energies $\approx 10-50$~KeV) that produce Type III bursts near $1$~AU \citep{krucker_etal_1999, haggerty_and_roelof_2002, haggerty_etal_2003}. 

Numerical simulations and associated theoretical formalisms for predicting the acceleration and transport of ion SEPs typically involve idealisations concerning one or more of the mean free path, scattering, magnetic field, dimensions, shock geometry, acceleration process, or imposed analytic approximations. For instance, the wave-particle scattering near the shock may be calculated with full time-dependent self-consistency \citep{ng_etal_2003} or assumed to proceed to completion with a steady-state diffusive shock acceleration solution \citep{lee_1983,lee_2005,zank_etal_2000,li_etal_2005,li_etal_2009,verkhoglyadova_etal_2010,vainio_etal_2014,hu_etal_2017}. Similarly a constant mean free path may be assumed for the particle transport \citep{Marsh2013,Marsh2015} or magnetic moment-induced focusing to small pitch-angles, magnetic-turbulence effects, and associated pitch-angle scattering included \citep{Jokipii66,matthaeus_etal_2003,zhang_etal_2009,shalchi_etal_2010,hu_etal_2017}, or cross-field diffusion \citep{zhang_etal_2009,droge_etal_2014,he_wan_2015}. Typically, the background magnetic field is assumed to be Parker-like and drift effects are ignored, but drift effects are sometimes included and found important \citep{Marsh2013,Marsh2015} and the magnetic field is sometimes significantly non-Parker-like \citep{schulte_etal_2011,schulte_etal_2012,li_etal_2016}. Finally, the formalisms available sometimes go beyond the usual one-dimensional approximation  to two dimensions \citep{kozarev_etal_2013}, or even three \citep{zhang_etal_2009,Marsh2013,Marsh2015}, and they can couple simulations of specific CMEs and their shocks with the particle transport formalism \citep{zank_etal_2000,verkhoglyadova_etal_2010,kozarev_etal_2013,hu_etal_2017}. Existing theory and simulations thus often experience major challenges explaining observed SEP events.

\section{Active Region Evolution, X-ray, Microwave, and Optical Flares}
\label{S3}

We begin the analysis of the 4 November 2015 solar eruptions with an overview of the source active regions, followed by the X-ray and microwave flare observations that define the initial stages of the three events. Table \ref{table_summary} provides a summary of these and other associated observations.  

\subsection{Source Active Regions}

\begin{figure}[htbp] 
\includegraphics[width=1.0\columnwidth]{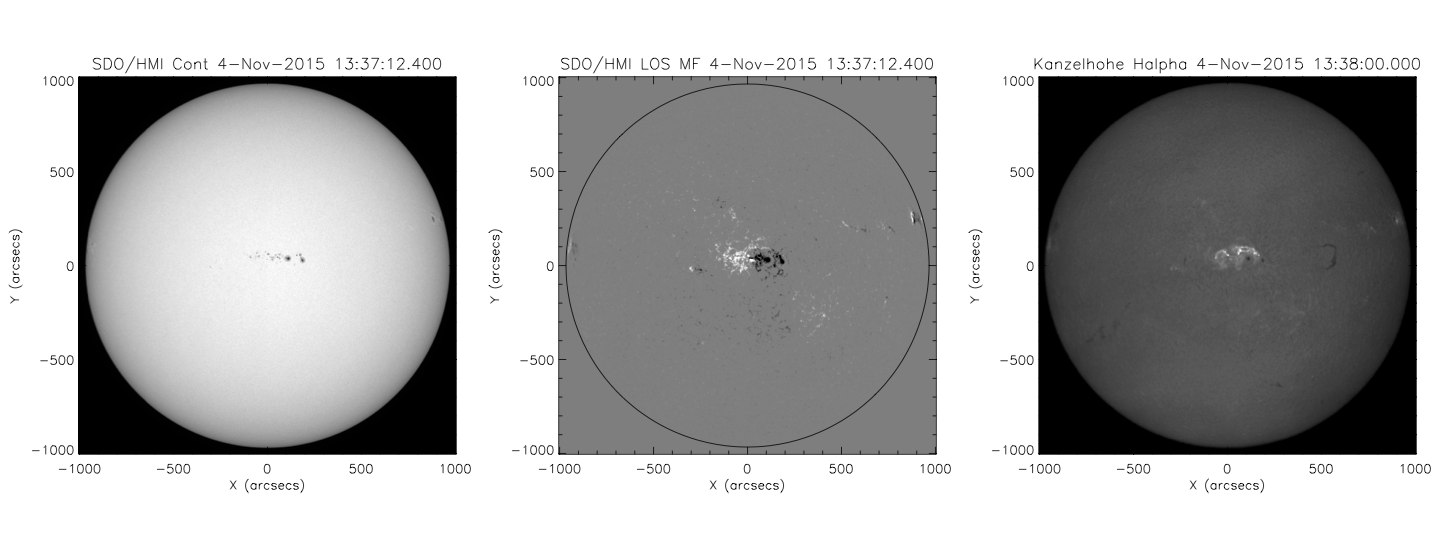}
\caption{Overview of the Sun at 13:37~UT on 4 November 2015. ({\it Left}) SDO/HMI continuum image, ({\it middle}) SDO/HMI line-of-sight magnetogram, ({\it right}) H$_{\alpha}$ image from the Kanzelh\"ohe Observatory, all recorded during the rise phase of Event~3.}
\label{Fig_SunOverview}
\end{figure}

Figure~\ref{Fig_SunOverview} overviews the Sun on 4 November 2015, showing a full-disk continuum image and a line-of-sight magnetogram from the {\it Helioseismic and Magnetic Imager} \citep[HMI;][]{Scherrer2012} onboard SDO together with an H$_{\alpha}$ filtergram from the Kanzelh\"ohe Observatory \citep{Poetzi2015}.  The two most prominent active regions present on the visible solar hemisphere on 4 November are NOAA AR 12443 located close to disk center (N6,W10), which is the source of Event~3, and AR 12445 located close to the western limb (N16,W76), which is the source of Events~1 and 2. AR12445 emerged and evolved very fast, over a period of four days, whereas AR 12443 was a long-lived active region. 

\begin{figure}[htbp] 
\centering
\includegraphics[width=1.0\columnwidth]{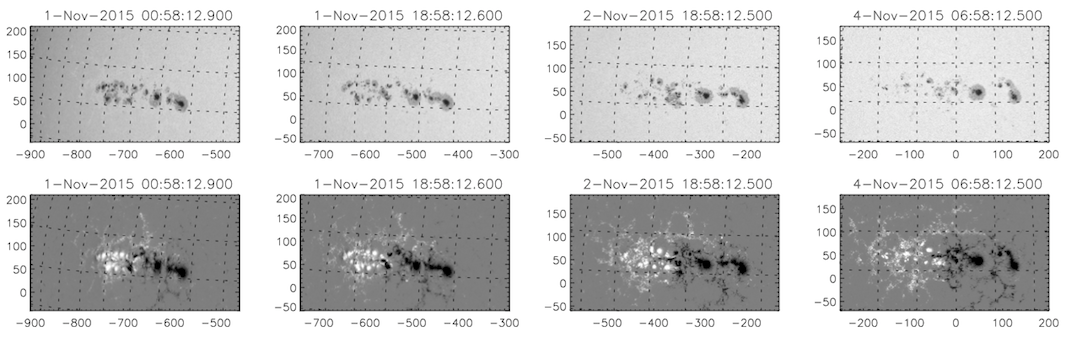} 
\caption{Evolution of NOAA AR 12443 from 1 to 4 November 2015. ({\it Top}) SDO/HMI continuum images and ({\it bottom}) SDO/HMI line-of-sight magnetograms (with the grayscale saturated at $\pm 1000$~G).}
\label{Fig_AR12443}
\end{figure}

\begin{figure}[htbp] 
\centering
\includegraphics[width=1.0\columnwidth]{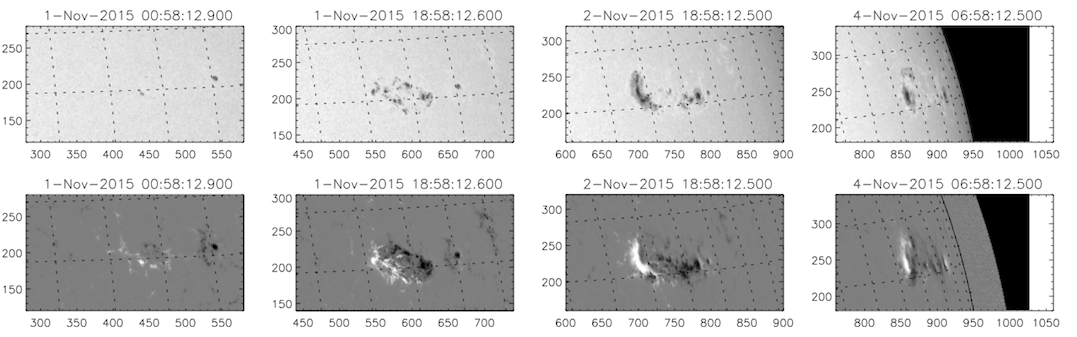} 
\caption{Evolution of NOAA AR 12445 from 1 to 4 November 2015. (Top) SDO/HMI continuum images and (bottom) SDO/HMI line-of-sight magnetograms (with the grayscale saturated at $\pm 1000$~G).}
\label{Fig_AR12445}
\end{figure}

NOAA AR~12443 is an extended AR region of McIntosh class Fck and magnetic Hale class $\beta\delta$ on 4 November. It developed when it was on the back side of the Sun and rotated onto the visible solar hemisphere on 28 October 2015. Figure~\ref{Fig_AR12443} shows snapshots of the evolution of NOAA AR 12443 on four days before 4 November when it produced Event~3 of our study. In contrast, AR~12445 developed very quickly. Figure~\ref{Fig_AR12445} shows the evolution of NOAA AR 12445 from 1 November when it was first visible and its fast flux emergence and development until 4 November, when it produced Events~1 and 2. On 4 November, its McIntosh class was Ekc and Hale class $\beta\delta$.

\subsection{X-ray and Microwave Emission for the Three Flares on 4 November 2015 -- Flare-related Electron Acceleration and Escape}

The microwave, soft, and hard X-ray emissions (SXR and HXR, respectively), associated with the three flares on 4 November 2015, are signatures and indicators of the electron-heating and acceleration processes occurring in the flaring active regions. HXRs at photon energies above about $20$~KeV are dominantly bremsstrahlung from nonthermal electrons interacting with the dense low corona and chromosphere. Microwaves, that is radio emission at frequencies between 1~GHz and several tens of GHz, are usually attributed to gyrosynchrotron radiation of electrons with energies between about $100$~KeV and a few MeV. It is necessary to look at the behavior of the spectrum in order to identify potentially competing processes: weak (flux densities below 100 sfu), slowly evolving bursts can also be due to thermal bremsstrahlung, and emission up to a few GHz to collective plasma processes. 

\subsubsection{Event 1}
\begin{figure}[htbp] 
\centering
\includegraphics[width=0.9\columnwidth]{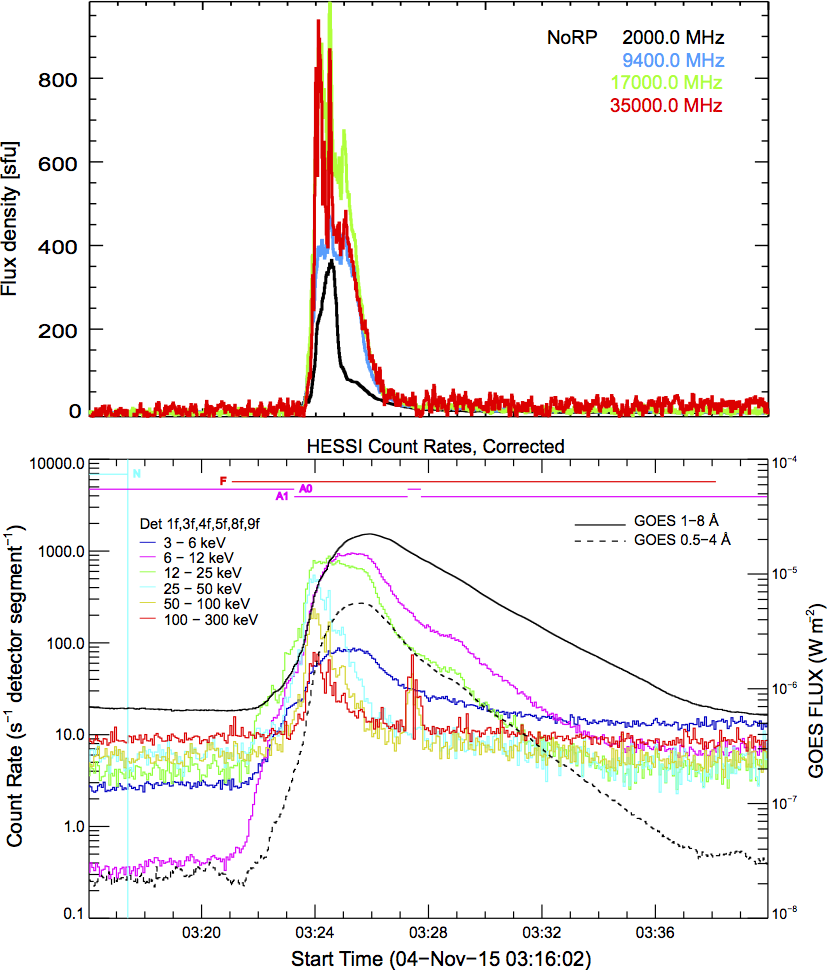}
\caption{Temporal histories of the NoRP microwave ({\it top panel}), RHESSI HXR, and GOES SXR ({\it bottom panel}) emission during Event~1.}
\label{Fig_MW1}
\end{figure}

Event 1 occurred at heliographic position (N15,W64) and reached  GOES class M1.9 (GOES start time: 03:20~UT, peak time: 03:26~UT). The time histories of the RHESSI {\it Hard X-Ray} (HXR), GOES {\it Soft X-Ray} (SXR) and the microwave emission from the {\it Nobeyama Radio Polarimeters} (NoRP) of the first flare are displayed in Figure~\ref{Fig_MW1}. They show an impulsive HXR and microwave burst during the rise phase of the SXR burst. The RHESSI HXR burst is observed to high energies, up to about 500~KeV, with the peaks near 03:24~UT above $25$~KeV and the $3 -- 12$~KeV channels peaking near 03:25~UT.  The microwave flux density spectrum has its maximum between 17 and 35~GHz, with a peak flux density of about $950$~sfu near 03:24 -- 03:25~UT. Although not exceptionally high, the flux density is well above values that can be achieved by thermal bremsstrahlung. The burst is hence due to gyrosynchrotron emission. The combination of a high peak frequency and moderately high flux density suggests that the emission occurs in a compact source with rather strong magnetic field, presumably at low coronal altitudes. The strong HXR and microwave emissions show that the parent flare is a very efficient accelerator of electrons to near-relativistic energies. 

\begin{figure}[htbp] 
\centering
\includegraphics[width=1.0\columnwidth]{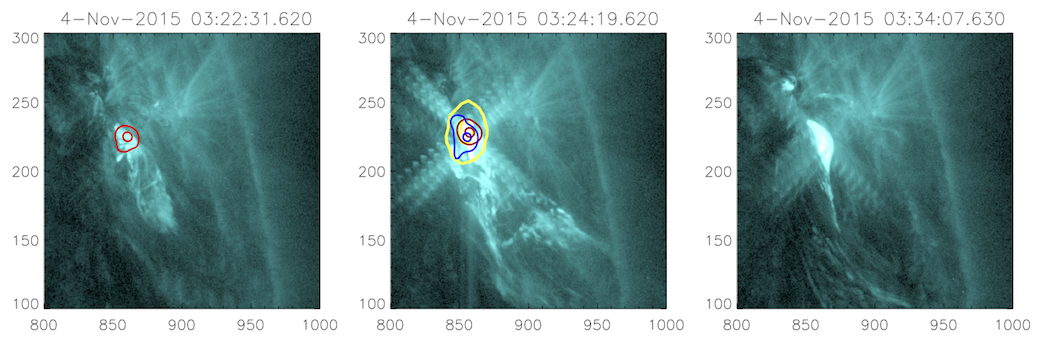}
\caption{Snapshots showing the evolution of Event 1 in AIA $131$~{\AA} filtergrams. Red and blue lines are contours from RHESSI X-ray images reconstructed during the early rise phase ({\it left}) and peak ({\it middle}) in the $6--12$ and $30--100$~KeV energy bands, respectively. The {\it rightmost image} is after the event. Yellow contours are from a NoRH $17$~GHz image at event peak. Units are arcseconds from Sun center.}
\label{Fig_aia1}
\end{figure}

Figure~\ref{Fig_aia1} shows snapshots of Event 1 observed by the SDO / Atmospheric Imaging Assembly (AIA) using the $131$~{\AA} EUV filter, sensitive to hot flaring plasma at temperatures of about 10~MK. The three images shown are recorded during the early rise, the peak and decay phase of the event. We overplot RHESSI $6--12$ and $30--100$~KeV sources as well as a $17$~GHz microwave image synthesized during the peak of the event from the Nobeyama Radioheliograph (NoRH). The RHESSI images have been reconstructed with detectors $2$ to $8$ \citep{Lin2002, Hurford2002}, but even with the fine grids included we are not able to resolve the emission from the flare loop and footpoints. All three instruments show that the flare is very compact, with the flare emission originating from a small loop arcade. The endpoints of the loops coincide with flare kernels observed in AIA $1700$~{\AA}. From the RHESSI and AIA images we estimate the distance between the loop footpoints to be about $20$~Mm, and the loop height $< 10$~Mm. These characteristics support the interpretation of the high peak microwave frequency observed by NoRP being due to the microwaves originating at low coronal altitudes in a compact source. The AIA $131$~{\AA} sequence plotted in Figure~\ref{Fig_aia1} also shows the impulsive filament eruption towards the South that also originates from the flaring region.

\begin{landscape}
\begin{table}
\begin{adjustwidth}{-3cm}{}
\caption{Short summary of the phenomena associated with the events of 4 November 2015.}
\begin{tabular}{c|c|c|c|c|c|c|c|c|c|c|c|c}     
\hline
& AR & SXR  & GOES  & & & & & EUV & CME & Shock /  & Space & SEPs \\
Event \# & Location & start time & X-ray & HXR & GHz & Type IIs & Type IIIs & wave speed & speed & CME & weather & \\ 
& & [UT] & class & & & & & [km s$^{-1}$] & [km s$^{-1}$] & at Earth & & \\
\hline
1 & 12445 & 03:20 & M1.9 & impulsive & Yes & Metric & Weak metric & 750 & 328 & No / No & No & Electrons \\
 & N15W64 & & & & No IP & IP & & S-SE dir. & PA 280\degree & & &No ions\\
& & & & & & & & & & & &\\
& & & & & & & & & & & &\\
2 & 12445 & 11:55 & M2.6 & short \& & No & Metric & No metric & 600 & 252 & No / No & No & No electrons \\
 & N12W73 & & & weak & No IP & No IP & & S-SE dir.  & PA 280\degree & & & No ions \\
& & & & & & & & & & & &\\
& & & & & & & & & & & &\\
3 & 12443 & 13:31 & M3.7 & strong \& & Yes & Metric & Metric & 700 & 580 & Yes / Yes & Yes & Electrons \\
& N09W04 & & & hard & IP & IP & & N-NW dir.  & Halo & & & Ions \\
\hline
\end{tabular}
\label{table_summary}
\end{adjustwidth}
\end{table}
\end{landscape}

\subsubsection{Event 2}
\begin{figure}[htbp] 
\centering
\includegraphics[width=1.0\columnwidth]{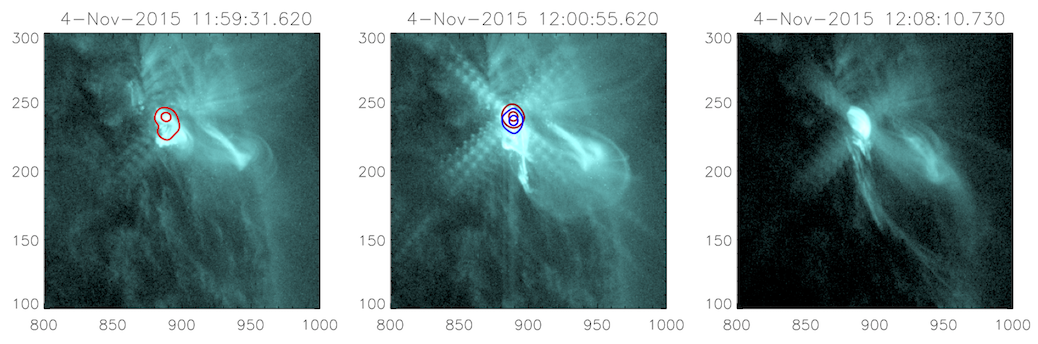} 
\caption{Snapshots showing the evolution of Event 2 in AIA $131$~{\AA} filtergrams. Red and blue lines are contours from RHESSI X-ray images reconstructed during the early rise phase ({\it left}) and peak ({\it  middle}) in the $6--12$ and $20--50$~KeV energy bands, respectively. The {\it rightmost image} is after the event.}
\label{Fig_aia2}
\end{figure}

The RHESSI HXR time profiles of the two later events are shown in Figure~\ref{Fig_HXR2-3}. Event 2 occurred at heliographic position N12W73  and reached GOES class M2.6 at (GOES start time: 11:55~UT, peak time: 12:03~UT). RHESSI covered the full impulsive phase of Event 2 and observed enhanced HXR emission up to energies of about $50$~KeV. The HXR emission of this event is clearly weaker and softer than the first event. However, the AIA $131$~{\AA} and RHESSI images plotted in Figure~\ref{Fig_aia2} reveal that Event~2 is  homologous with Event~1, as regards the occurrence in the same region, the small and compact flare loops  (also reflected in the short HXR and SXR emission profiles), and the associated ejection of filament material toward the south. At microwave frequencies RSTN ({\it Radio Solar Telescope Network}, operated by the US Air Force) sees a weak burst of gyrosynchrotron emission with peak frequency $8.8$~GHz, between 12:00 and 12:04 UT. Despite the homology, Event 2 is a much less efficient electron accelerator than Event 1 based on the X-ray and microwave emissions.

\subsubsection{Event 3}
\begin{figure}[htbp] 
\centering
\includegraphics[width=1.0\columnwidth]{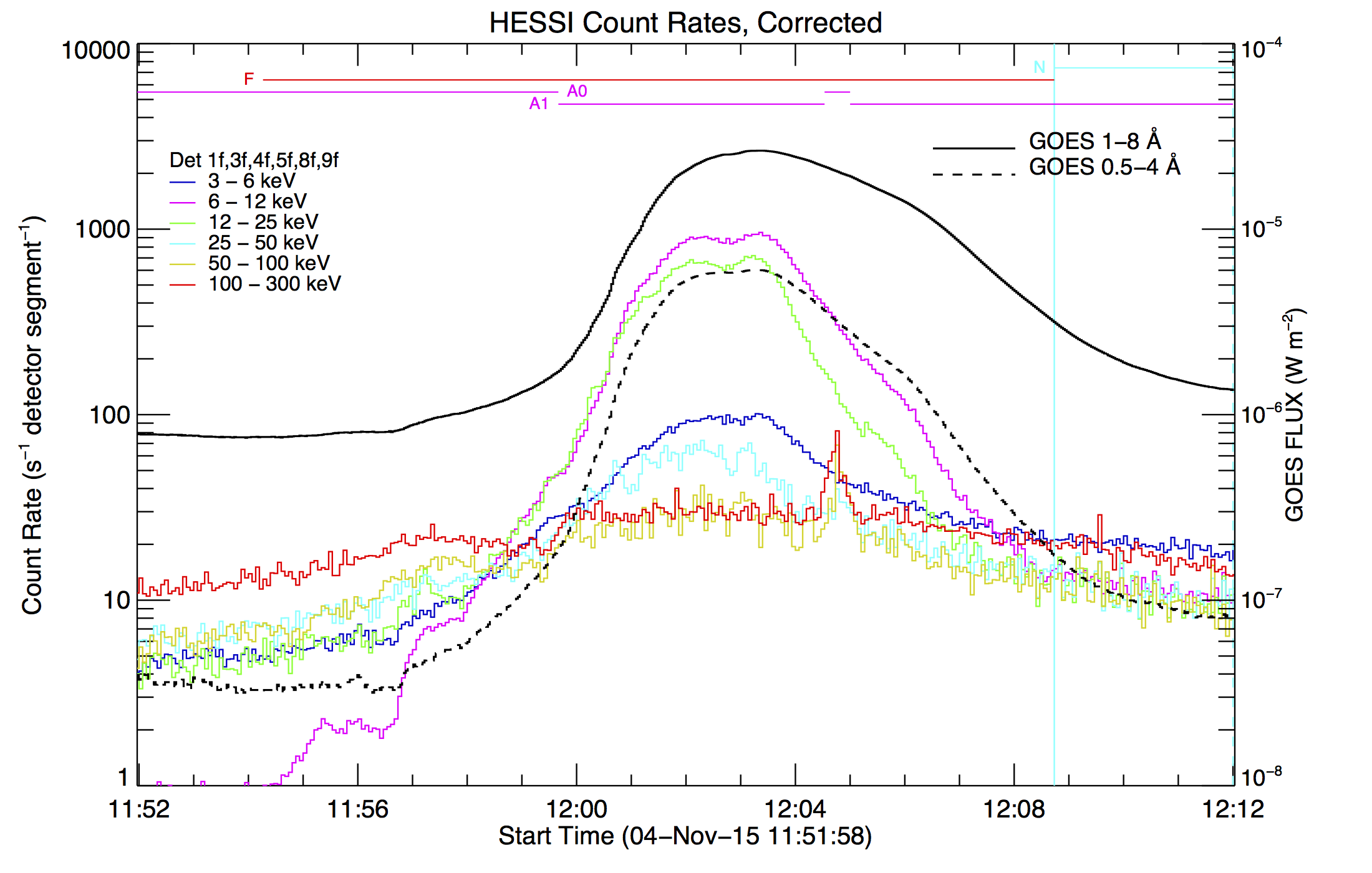} 
\includegraphics[width=1.0\columnwidth]{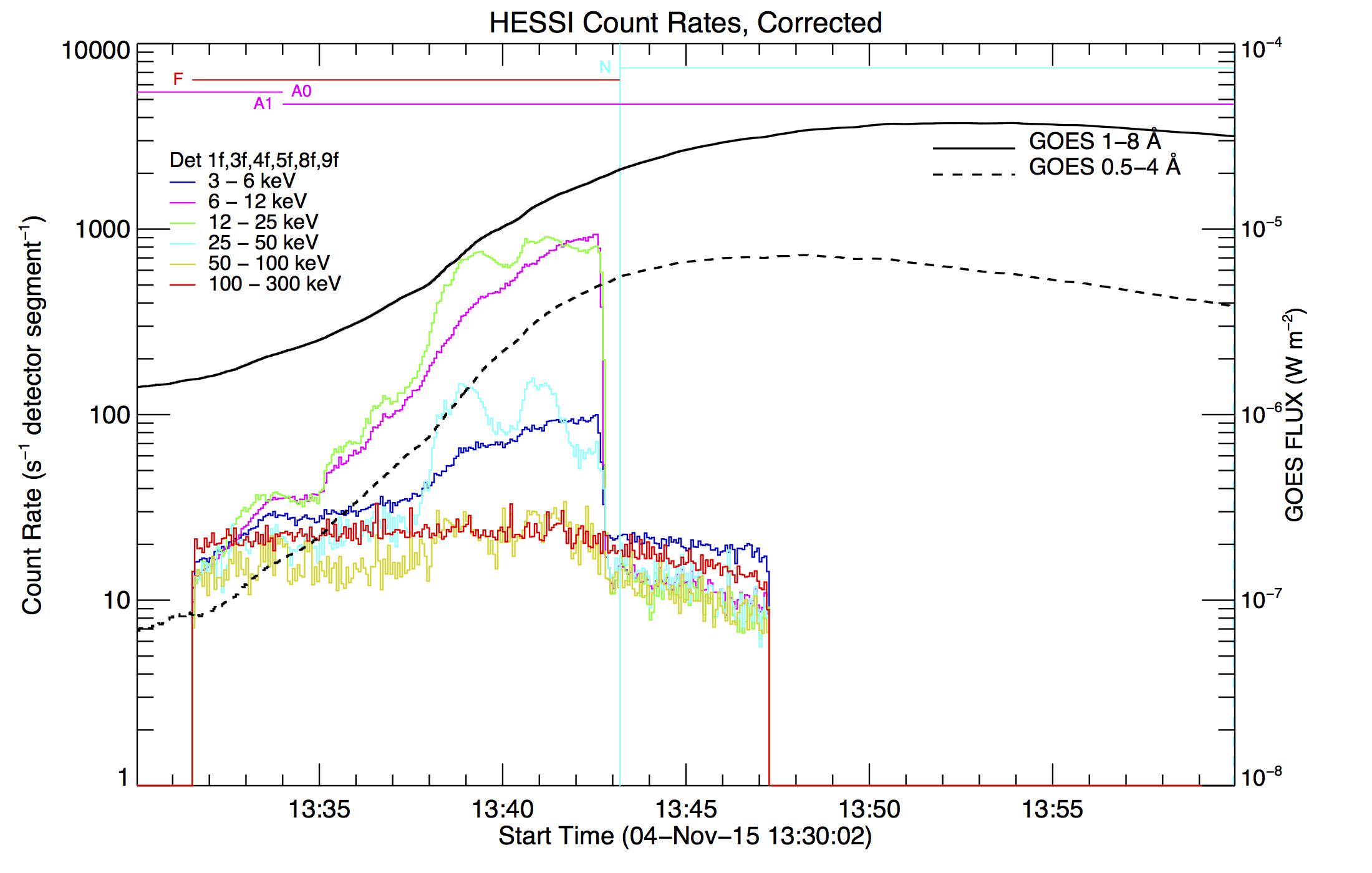} 
\caption{Time histories of the RHESSI HXR emission during Event 2 ({\it top}) and Event 3 ({\it bottom}).}
\label{Fig_HXR2-3}
\end{figure}

\begin{figure}[htbp] 
\centering
\includegraphics[width=1.0\columnwidth]{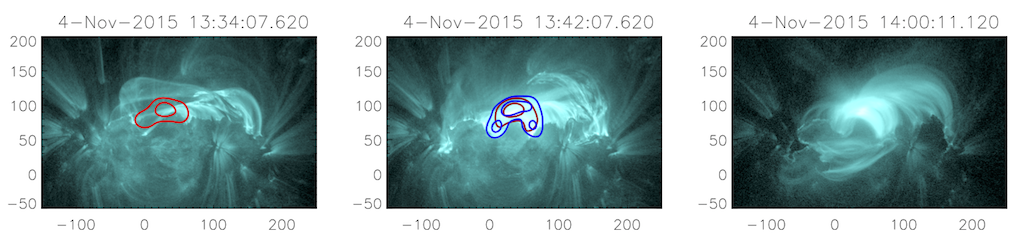} 
\caption{Snapshots showing the evolution of Event 3 in AIA $131$~{\AA} filtergrams. Red and blue lines are contours from RHESSI X-ray images reconstructed during the early rise phase ({\it left}) and peak ({\it middle}) in the $6--12$ and $20--50$~KeV energy bands, respectively. The {\it rightmost image} is after the event.}
\label{Fig_aia3}
\end{figure}

Event 3 occurred close to disk center, at heliographic position (N09,W04). It is the largest of the three events under study, with a GOES class M3.7 (start time: 13:31~UT, peak time: 13:52~UT). GOES and RHESSI light curves are shown in the bottom panel of Figure~\ref{Fig_HXR2-3}. Classifying the GOES light curves in Figures \ref{Fig_MW1} and  \ref{Fig_HXR2-3} using the system of \citet{cane_etal_1986}, Events 1 and 2 are impulsive events and Event 3 a gradual event. The RHESSI hard X-ray observations are restricted to before 13:43~UT due to spacecraft night. However, a comparison with the light curves from {\it FERMI}/GBM shows that RHESSI missed no major burst. The HXR burst is observed up to photon energies of about $100$~KeV. The radio emission, however, does show efficient electron acceleration, starting with a moderately strong burst with peak frequency near $9$~GHz in the impulsive phase ($\approx$ 13:38--13:50), and followed by a long-duration burst in the post-impulsive phase (14:00--15:00 UT; Type IV burst) that was mainly observed at frequencies below 3 GHz (cf. Section \ref{S5}). Figure~\ref{Fig_aia3} shows snapshots during the early rise, peak, and decay phase of Event~3 in AIA~131~{\AA} together with RHESSI HXRs. In contrast to the compact Events 1 and 2, Event 3 shows an extended flare arcade. The East-West extent of the overall flaring region as observed in the AIA EUV emission is about $140$~Mm. The RHESSI emission is concentrated mostly to the brightest flaring loops observed in AIA $131$~{\AA} (compare the middle and right panels), with a loop footpoint separation of about $50$~Mm, corresponding to a loop apex height of $25$~Mm for a semicircular loop.

\section{Failed Filament Eruptions, EUV Waves, and CMEs}
\label{S4}

\subsection{Filament Eruptions and EUV Waves}

\begin{figure}[t!]
\centering
\includegraphics[width=1.0\columnwidth]{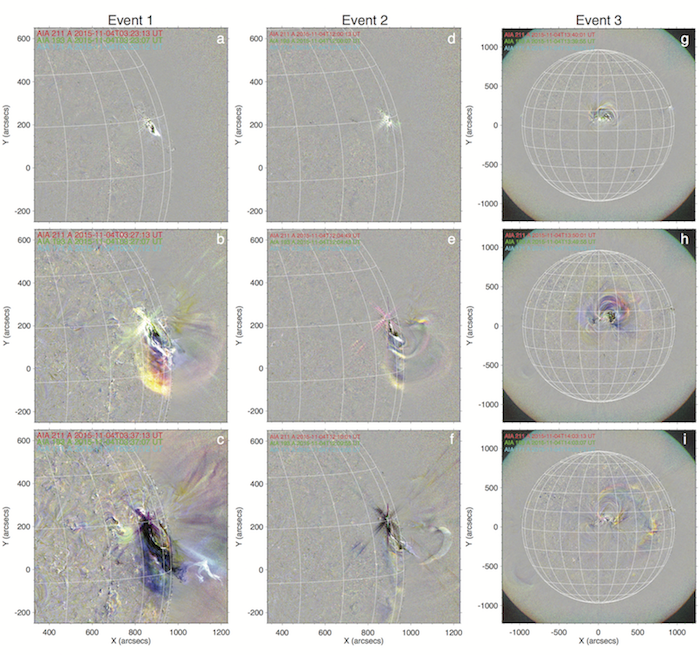} 
\caption{Summary of eruption evolution for the three events occurring on 4 November 2015. ({\bf a})--({\bf c}) Event 1, beginning with a compact flare-brightening at (N14,W65) at 03:23~UT, followed by a filament eruption and a EUV front which largely propagate in a southwesterly direction. The filament eruption partly failed, with material falling back to the surface and erupting outwards. ({\bf d})--({\bf f}) Event 2 has similar evolutionary characteristics to Event 1, beginning with a flare-brightening from the same active region and a filament eruption. The filament is surrounded by two distinct yellow structures, the outer one being an EUV front, while the inner one develops into a loop-like feature seen in panel (f). The filament eruption is smaller than Event 1 and largely failed. ({\bf g})--({\bf i}) Event 3 occurs at disk center, beginning with a brightening of coronal loops and the propagation of a diffuse EUV front in the northwest direction.}
\label{fig:eruption_summary}
\end{figure}

Figure~\ref{fig:eruption_summary} summarises the eruption evolution seen in EUV for the three events, clearly showing moving material and propagating wave features in all three cases. Red, green, and blue colors correspond to AIA data number values for the filters $171$, $193$, and $211$~\AA, respectively. Events 1 and 2 show very similar morphology eruptions off the west limb, generally directed towards the south, while Event 3 shows the event at disk center, beginning with a brightening of coronal loops, followed by the propagation of a diffuse front, largely in northwesterly direction out to $\approx 50$\degree~W longitude, before decreasing in intensity.

\begin{figure}[t!]
    \centering
    \includegraphics[width=1.0\columnwidth]{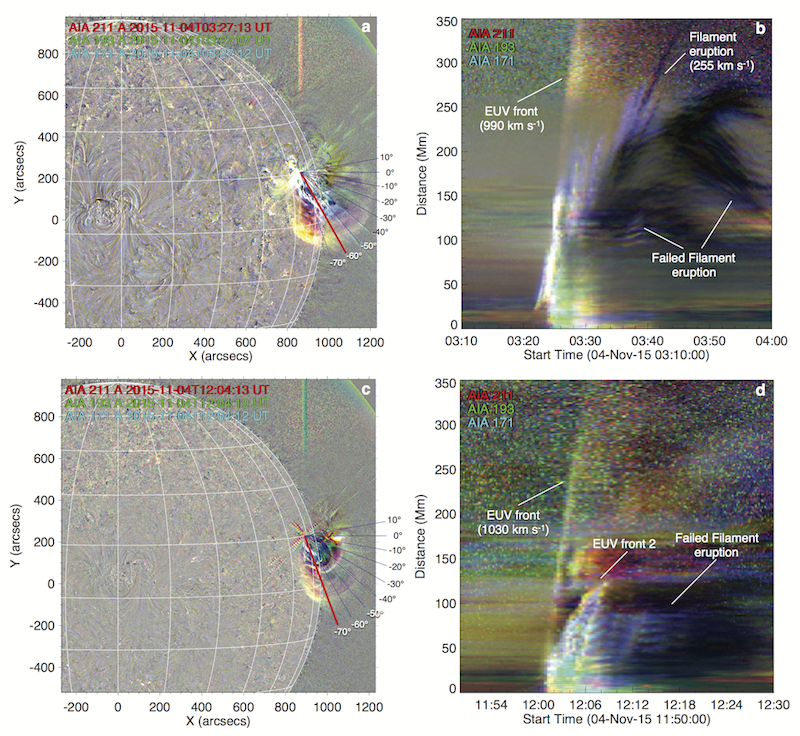}
    \caption{EUV kinematics summary for Events 1 and 2. ({\bf a}) Snapshot showing the EUV structures. The blue lines that originate at the erupting active region show the directions along which distance-time maps were produced. ({\bf b}) Distance--time map taken along the red line traced at angle $-60$\degree in panel ({\bf a}), showing the EUV front, filament eruption, and the failed section of the filament eruption. ({\bf c}) Event 2 eruption in the same format as ({\bf a}). The line highlighted in red is the trace ($-70$)\degree along which the distance--time map shown in panel ({\bf d}) was constructed. It shows several erupting features including the outer EUV front and the erupting loop.}
    \label{fig:dt_analysis}
\end{figure}

For Events 1 and 2, most of the intensity appeared off-limb. In order to analyse the off-limb kinematics of Event 1, we extracted intensity traces from lines at several angles originating from the active region as indicated in Figure~\ref{fig:dt_analysis}a. For one angle, intensity traces at successive times were stacked to produce a distance--time map. The distance-time maps were made for the $171$, $193$, and $211$~\AA~ filters and then combined in the usual RGB format so activity at the three wavelengths may be represented simultaneously in a single map. Panel b shows one such distance-time map for the $-60$\degree trace  indicated by the red line in Panel a. The map reveals that the eruption along this direction was composed of several distinct features, including a bright yellow EUV front that becomes visible at 03:22~UT and reaches a speed of $990$~km~s$^{-1}$; this was followed by escaping filamentary material with a speed of $255$~km~s$^{-1}$ and finally the failed filamentary eruption which starts to fall back to the solar surface at 03:45~UT. Traces taken along angles from $-60$\degree to $-30$\degree show similar behaviour, while traces along angles from $-20$\degree to $0$\degree only show the bright yellow front (no following filament), propagating at speeds of $\sim 800$~km~s$^{-1}$ in this direction. No radially propagating feature could be traced in the $10$\degree trace or above this angle.

The distance--time analysis performed for Event 2 is shown in Figure \ref{fig:dt_analysis}c and d, where the example distance--time map is from the $-70$\degree trace as indicated in Panel c. Again several eruption-related features may be identified, the fastest of which is a bright EUV front that begins just after 12:00~UT and propagates with a speed of $1030$~km~s$^{-1}$. This is followed by a much slower front (most likely the feature which develops into a coronal loop in the images) and then non-erupting filamentary material -- the failed filament eruption is not as pronounced as Event 1. The same features may be identified in the distance--time maps that are oriented towards the south, {\it i.e.}, $-70$\degree to $-30$\degree, while traces at $-20$\degree to $+10$\degree show only the slow secondary yellow front (loop) propagating at speeds of $\approx 350$~km~s$^{-1}$.

\begin{figure}[t!]
    \centering
    \includegraphics[width=1.0\columnwidth]{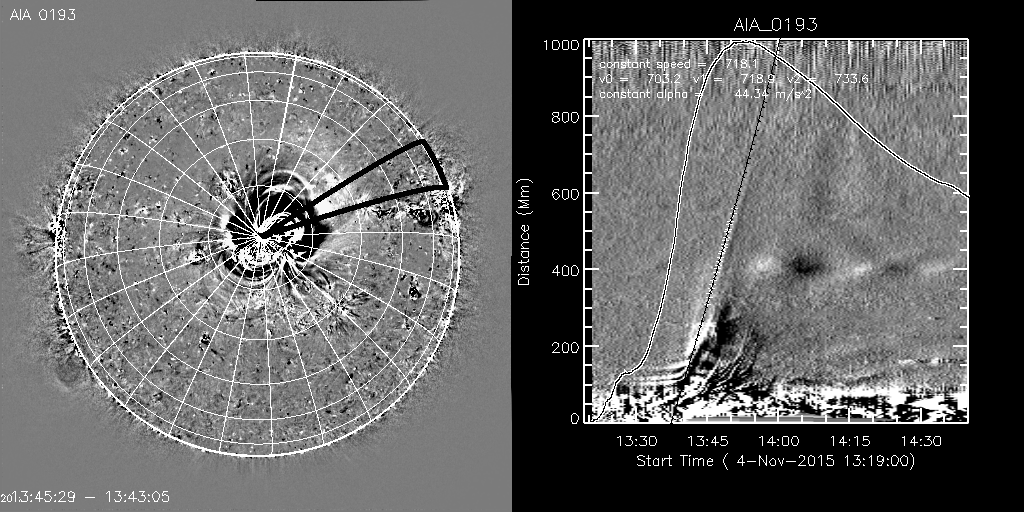}
    \caption{The kinematics of the EUV wave for Event 3. ({\it Left}): The speed and acceleration are calculated along great circles averaged in $24$ $15$\degree-wide sectors that originate at the flare center. ({\it Right}): Space--time plot of the sector marked in black in the left panel. The normalized GOES $1-8$~\AA\ light curve is overplotted using a black line superposed on a broader white curve. Even in running-- difference images the wave front is diffuse.  The selected sector is one of the few that let us trace the front edge of the wave.}
    \label{event3_kinematics}
\end{figure}

EUV waves from Events 1 and 2 were also seen to propagate on the disk, limited to the south to southeast directions. The speed of the wave as it propagated on the disk was measured along the great circle passing the flare \citep{nitta_et_al_2013}. The speed was lower on the disk than that measured off-the-limb. It was $\approx 750$~km~s$^{-1}$ and $\approx  600$~km~s$^{-1}$, respectively, for Events 1 and 2. The EUV wave from Event 3 was diffuse and anisotropic, identifiable only westwards of  AR~12443 with the brightest part moving northwestward (Figure~\ref{event3_kinematics}). The speed in that direction was $\approx 700$~km~s$^{-1}$. The EUV wave was seen between 13:40~UT and 13:50~UT, comparable to the period in which Type II bursts were observed (see Section~5). A more detailed analysis of this EUV wave will be given elsewhere.

\begin{figure}[t!]
\centering
\includegraphics[width=1.0\columnwidth]{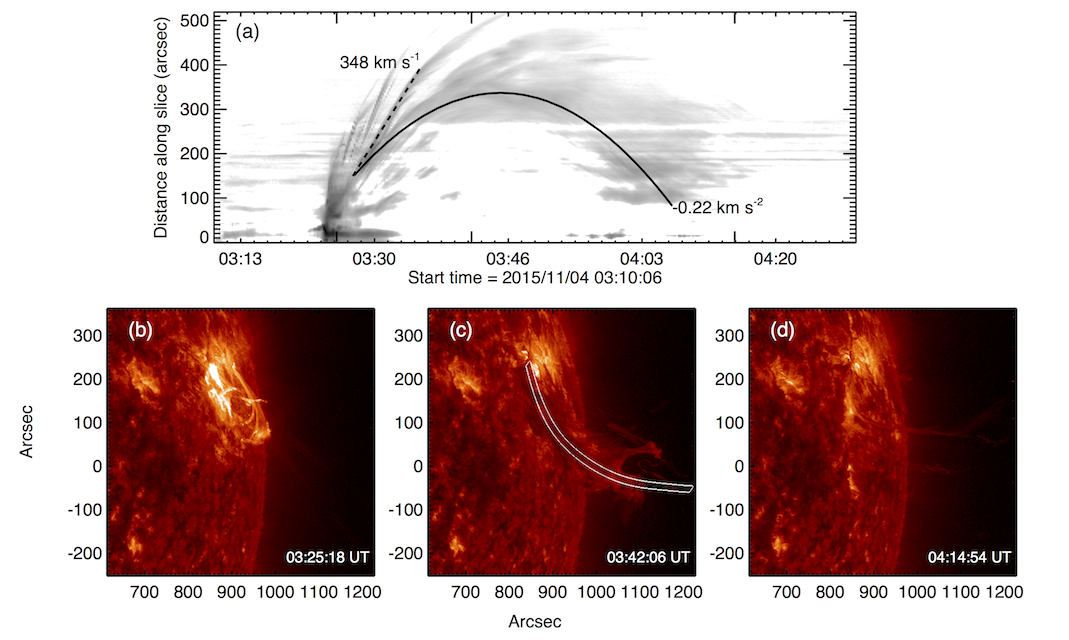}
\caption{Filament eruption kinematics from $304$~\AA\ SDO/AIA observations of Event 1. ({\bf a}) Background-subtracted distance-time plot using the curved trajectory in ({\bf c}). ({\bf b})--({\bf d}) Snapshot $304$~\AA\ SDO/AIA images at three times. The filament is ejected behind and somewhat slower than the EUV wave, and the failed eruption material returns at slightly lower than gravitational acceleration ($0.27$~km~s$^{-2}$).}
\label{fig:eruption_patrick}
\end{figure}

The early stages of the failed eruptions in the low corona were analysed using SDO/AIA for Events 1 and 2, in order to better understand the CMEs they drove. Figure \ref{fig:eruption_patrick} shows the failed filament eruption for Event 1, using 304~\AA~data from SDO/AIA. The compact flare of Event 1 was followed by a filament eruption toward the Southwest in the plane of the sky, along a curved path that is initially at a small angle to the local chromospheric surface. An EUV front and a weak Moreton wave (identified in movies of H$_{\alpha}$ filtergrams from GONG and Kanzelh\"ohe Observatory) accompanied the eruption, propagating both southwards on-disk and in a southwesterly direction off-limb. By 03:37~UT the EUV front had left the AIA field of view and started to dissipate in intensity. While some of the filament was completely ejected, a large portion of it fell back to the solar surface, both towards the active region, and to an area south of the active region. The falling material produced bursts of $304$~\AA~radiation when it hit the chromosphere, presumably from the heating of the plasma.

The motion of the filament material can be analysed quantitatively by studying the location (or distance) of bright or dark features as a function of time along a specified path \citep{McCauley:2015}.  Taking into account the curved path followed by the erupting material in the bottom panels of Figure \ref{fig:eruption_patrick} leads to the distance--time diagram in the top panel.  The linear segment of one fast feature in the top panel corresponds to a speed of $348$~km~s$^{-1}$, with others having top speeds of $\approx 500$~km~s$^{-1}$. The downwards curvature corresponds to acceleration sunwards. The parabolic line corresponds to an acceleration of $0.22$~km~s$^{-2}$, agreeing very closely with the Sun's predicted gravitational acceleration (of $0.27$~km~s$^{-2}$) at this distance.  Note that there is a difficulty understanding the paths taken by the erupting and falling material: if the curved eruption path is interpreted as following the local magnetic field, then the more vertical path taken by the falling matter suggests either that the field direction has changed substantially or that the falling matter does not follow the field lines.

\subsection{CMEs for Events 1 and 2}

\begin{figure}[htbp] 
   \centering
   \includegraphics[width=1.0\columnwidth]{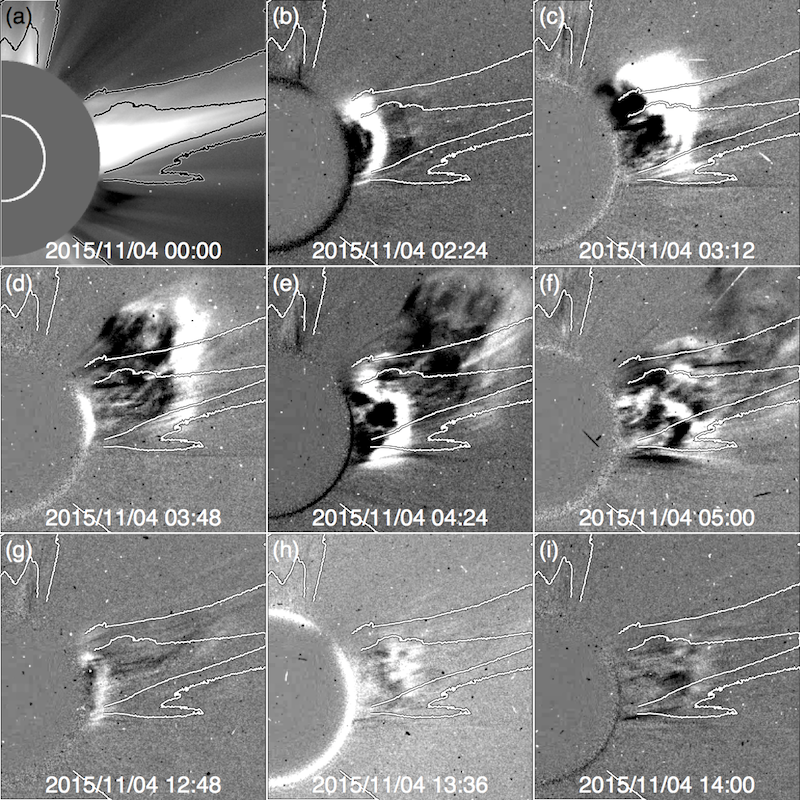}  
   \caption{The evolution of the CMEs for Events 1 and 2, as seen by the LASCO instrument's C2 camera onboard SOHO. See text for details.}
   \label{figure_c2}
\end{figure}

Figure~\ref{figure_c2} shows the evolution of the first two CMEs in the field of view of the C2 camera ($2 -- 6 \rsun$) of the {\it Large Angle and Spectrometric Coronagraph} \citep[LASCO]{brueckner_etal_1995} on board SOHO. Figure~\ref{figure_c2}a is a direct image from LASCO~C2 taken at 00:00~UT on 4 November, where the black contours identify the location of the coronal streamers shaping the pre-event topology of the solar corona. Figures~\ref{figure_c2}b-i are difference images at the indicated times and the white contours indicate the location of the coronal streamers identified in Figure~\ref{figure_c2}a. Figures~\ref{figure_c2}b--c at 02:24~UT and 03:12~UT show an unrelated CME (first seen in LASCO~C2 at 02:00~UT) that occurred prior to the events under study. This preceding CME came from a small sigmoid eruption from the decaying AR 12441, which was downgraded to a non-numbered region on 4 November.    

Figures~\ref{figure_c2}d--f (second row) show the evolution of the CME associated with Event 1, which was first seen in C2 at 03:48~UT. It propagated mostly within the streamer, and by ~05:00~UT it broke into different parts. In fact it is extremely difficult to see it in the field of view of LASCO~C3 ($3.5-30 \rsun$). The weak and fragmented appearance of this CME may be due to its propagation within the coronal streamer.  The estimated plane-of-sky speed at a position angle of $280$\degree ({\it i.e.} $10$\degree above the equatorial plane) was $328 \pm 8$~km s$^{-1}$. 

Similarly, Figures~\ref{figure_c2}g--i (third row) show the evolution of the CME associated with Event 2. This CME was first seen in the LASCO/C2 field of view at 12:36~UT. It propagated within the streamer and by ~14:36~UT, just before the CME from AR 12443 associated with the Event 3 occurred, it was already very difficult to see in the LASCO~C2 images. The estimated plane-of-sky speed at a position angle of 280\degree ({\it i.e.} $10$\degree above the equatorial plane) was $252\pm$14~km s$^{-1}$.

Although the CMEs associated with Events 1 and 2 were initially impulsive without a clear driver, the magnetic structure in which they propagated ({\it i.e.} the coronal streamer) might have played a role in their fast weakening, ragged structure, and rapid decay. Similarly, the generation of the metric Type II emissions observed (see Section~5) may have been favored by the closed magnetic field structure at the base of coronal streamer encountered by these two weak CMEs \citep[{\it e.g.}][]{kong_etal_2015}. As shown below, the fast decay of the CMEs did not favor the production of interplanetary Type II emissions.

\subsection{CME for Event 3}

\begin{figure}[h] 
   \centering
   \includegraphics[width=1.0\columnwidth]{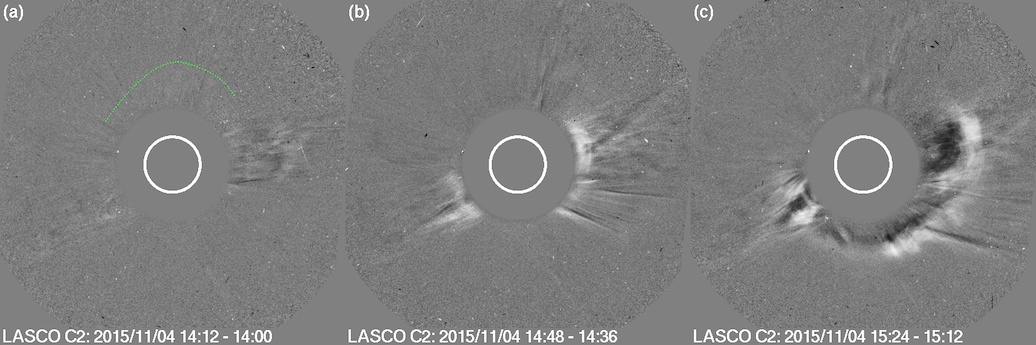}  
   \caption{The evolution of the CME for Event 3. See text for detail.}
   \label{cme_event3}
\end{figure}

According to the CDAW LASCO CME catalog (\url{http://cdaw.gsfc.nasa.gov/CME\_list}), the CME associated with Event 3 first appeared at 14:48 UT in the West to Northwest directions and then  developed into a full halo CME; see Figures \ref{cme_event3}b and c, respectively. Its average plane-of-the-sky speed in the LASCO field of view was $\approx 580$~km~s$^{-1}$ with a slight acceleration that led to a speed of $\approx 620$~km~s$^{-1}$ at the last measurable height of $15 \rsun$. Note that the CME front is far from smooth, suggestive of multiple flux ropes or several distinct erupting structures contributing to it. Close examination of the LASCO difference movies reveals a very diffuse front, indicated by a green curve in Figure \ref{cme_event3}a, that moves northward between 14:00 UT and 14:24 UT. The average speed in the image plane was $\approx 1200$~km~s$^{-1}$.

\section{Coronal and Interplanetary Radio Bursts}
\label{S5}

\begin{landscape}
\begin{figure}[h]
\centering
\includegraphics[width=1.0\columnwidth,trim=3cm 0cm 0cm 2cm]{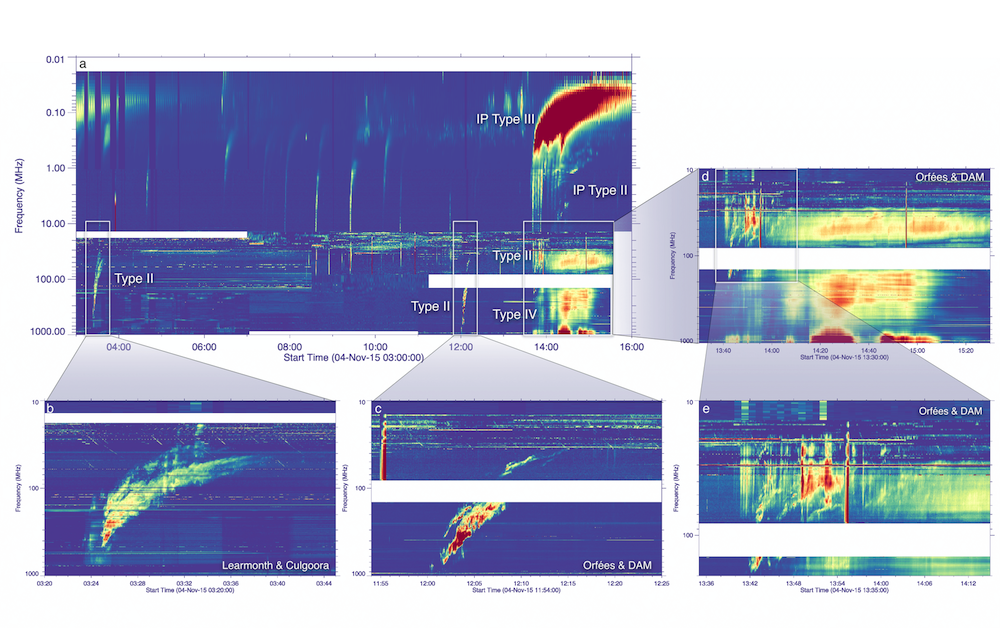}
\caption{({\bf a}) Summary of all radio dynamic spectra for the events on 4 November 2015. Constructed from a mosaic of {\it Wind}/WAVES, Learmonth, Culgoora, Orf\'{e}es, Nan\,cay (NDA), and Callisto (Bleien, Mauritius, and Gauribidanur). ({\bf b}) Event 1 was associated with a metric Type II radio burst that began at $\approx 600$~MHz that does not extend obviously below $10$~MHz. Weak Type III radio bursts occur prior to the Type II burst and then start at the same frequency as the Type II near 03:30:15, 03:31:15, and 03:33:00~UT. These are associated with a small interplanetary Type III observed by {\it Wind}/WAVES in ({\bf a}). ({\bf c}) Event 2 was associated with strong metric Type II emission, beginning at $\approx 800$~MHz. No interplanetary Type IIs or IIIs were observed for this event. ({\bf d})-({\bf e}) Event 3 was associated with at least three metric Type II bursts (see the details in ({\bf e}), multiple metric Type III bursts, and a Type IV burst over the band $20 - 700$~MHz. These metric events led to an interplanetary Type II and numerous interplanetary Type IIIs.}
\label{fig:radio_summary}
\end{figure}
\end{landscape}

Figure~\ref{fig:radio_summary} shows a summary of all radio bursts occurring on 4 November 2015 during the period 03:00--16:00~UT in the domain $0.01-1000$~MHz. The figure consists of a mosaic of plots from the Wind/WAVES, Learmonth, Culgoora, Orf\'{e}es, Nan\,cay Decametric Array (referred to as NDA below), and Callisto (Bleien, Mauritius, and Gauribidanur) spectrographs. 

From Figure~\ref{fig:radio_summary}a, the three events start near 03:25~UT, 12:00~UT and 13:40~UT. All three events have strong and complex metric Type II radio bursts, all with multiple lanes and both fundamental and harmonic emission. The metric Type III characteristics differ strongly between Events 1 and 2 on the one hand, and Event 3 on the other hand. Unlike the other two, Event 3 has a strong Type IV burst.

\subsection{Event 1}

Figure~\ref{fig:radio_summary}b is an enlargement of Event 1's dynamic spectrum using Learmonth and Culgoora data. It is complex, showing evidence for multiple lanes and several time-varying fundamental and harmonic bands. The strongest harmonic emission starts near $450$~MHz, an unusually high frequency,  at $\approx$ 03:25~UT and drifts to $\approx 50$~MHz over period of $\approx 12$ minutes, although evidence for weaker Type II-like emission exists near $800$~MHz near 03:24~UT. Weak fast-drifting signals from near 03:23:30 to near 03:25~UT at frequencies $\approx 100 - 800$~MHz are likely Type III bursts. They are the low-frequency counterpart of the impulsive microwave burst (Figure \ref{Fig_MW1}). These bursts are cut off near 100 MHz, with no counterpart in the high corona and interplanetary space. Similar Type III bursts start near 03:30:15, 03:31:15, and 03:33:00~UT close to the frequency of the harmonic Type II burst and drift to lower frequency. These latter bursts might be interpreted in terms of SA events \citep{cane_etal_1981, cane_etal_1984, bougeret_etal_1998, reiner_and_kaiser_1999}, where SA variously stands for Shock Accelerated \citep{cane_etal_1981, cane_etal_1984} or Shock Associated \citep{bougeret_etal_1998} or complex Type III-like \citep{reiner_and_kaiser_1999} events, as discussed further below. They are most likely related to the weak interplanetary Type III that is visible in expanded views of the {\it Wind}/WAVES data near these times in the restricted frequency range $8 -- 14$~MHz and then becomes clearly visible in the approximate frequency range $150 - 600$~KHz and period 03:40 - 03:50~UT. The fact that there is no strong Type III emission from coronal (RSTN) to interplanetary ({\it Wind}/WAVES) frequencies for this event, despite the evidence for efficient electron acceleration from X-ray and microwave observation (see Section \ref{S3}), suggests that either most of the accelerated electrons did not reach open magnetic field lines \citep{axisa_1974, Kle:al-10}, or that they were radio-quiet along open field lines \citep{li_and_cairns_2012, li_and_cairns_2013}. The observation of SEP electrons for Event 1 (see Section~7.1) provides strong evidence for the latter interpretation. The multiple lanes of Type II emission, possibly with band-splitting as well, indicate either i) that a single shock has multiple source regions with different densities simultaneously producing observable radio emission or ii) that more than one shock exists and produces radio emission in distinct regions with different densities. 

\subsection{Event 2}

Figure~\ref{fig:radio_summary}c shows Event 2's Type II radio burst, observed using Orf\'{e}es and NDA. Overall, this burst was much weaker than the other two. It is also complex, with evidence for multiple lanes that have different frequency-drift rates and sometimes overlap. The event has clear fundamental and harmonic structure, starting at the unusually high frequencies of $\approx 350$~MHz and $\approx 700$~MHz near 12:01:30~UT, and drifts to $30$~MHz over a period of 12~minutes. Interestingly, the Type II event starts with quite an intense harmonic that diminishes quickly. By the time the event enters the NDA frequency domain ($\approx 10 - 80$~MHz) it has faded to primarily a weak fundamental band, again with evidence for multiple lanes, and perhaps some very weak harmonic emission. There was no significant interplanetary Type II activity associated with this radio event or the faint microwave burst seen approximately 12:00 -- 12:04 UT (Section 3). There was also no discernible Type III emission at coronal or interplanetary frequencies, a small difference from the weak Type III emission for Event~1, again interpretable in terms of electrons not being released onto open field lines or not having the right properties to produce observable radio emission.

\subsection{Event 3}

Figures~\ref{fig:radio_summary}d and e show enlargements of the radio bursts of Event 3 above 10~MHz. This was by far the most complex of the three radio burst events. It begins with Type~III bursts starting shortly before the Type~II burst appears. These initial Type IIIs are easily seen near $1$ GHz and below $80$~MHz, while the first Type II emission begins near 13:42~UT and 200~MHz. This Type II mainly exists at NDA frequencies and shows very complex and sporadic bands of emission. Indeed an unusually large number of multiple lanes (not distinguishing between bands and split-bands) are identifiable (at least $6$ and perhaps up to $11$ or beyond depending on the observer's definition) and for an unusually long period, from 13:42~UT near $200$~MHz until about 14:08 UT near $30$~MHz. In addition a broadband, long-lasting, Type IV radio burst exists between about $30$ and $1000$~MHz, typically at higher frequencies than the Type II burst and extending long after the Type II burst has ceased. The Type IV burst is particularly intense at frequencies above 700~MHz, extends to unusually low and high frequencies, and appears to show strong vertically-aligned fine structures, particularly above the NDA domain for 13:50 \,--\, 13:56~UT and after about 14:00~UT. More detailed analysis is required to determine whether or not the fine structures show the increase of duration with decreasing frequency that characterises Type III bursts. The Type IV emission also shows pulsations in intensity on timescales of minutes. 

Type III bursts are also observed during much of Event 3, starting with a strong set in the period 13:40 \,--\, 13:42~UT before the Type II emission starts and then continuing intermittently before some more intense events occur at frequencies below the Type II burst during the period 13:49 \,--\, 13:54~UT. These latter events intensify at the frequency of the Type II burst, suggesting a physical association, but they may also have counterparts at frequencies above $\approx 150$~MHz in the Orf\'{e}es domain. 

\subsection{Interplanetary and {\it In-Situ} Observations}

Returning to Figure~\ref{fig:radio_summary}a for Event 3, the interplanetary extensions of the metric Type IIs are clearly present until at least 16:00~UT and a frequency of 500~KHz, but plausibly until 17:15~UT and 200~KHz. Fundamental and harmonic pairs are evident. Furthermore, multiple lanes are present again, with one restricted to above 4~MHz before about 14:30~UT and the limits on the other set described previously. 

Multiple distinct interplanetary Type IIIs are evident in the frequency range 400~kHz to 14~MHz of Figure~\ref{fig:radio_summary}a, as detailed more below, merging into an extended burst below 200~kHz that has little structure (with this color bar) and lasts until after 16:00~UT. This type of unusually long and bright group of Type III bursts is often called a Type III-L event \citep{cane_etal_2002}. Note that the two sets of metric Type IIIs discussed above connect to multiple individual interplanetary Type IIIs and the merged burst below $200$~kHz. 

\begin{figure}[h] 
\centering
\includegraphics[width=1.0\columnwidth]{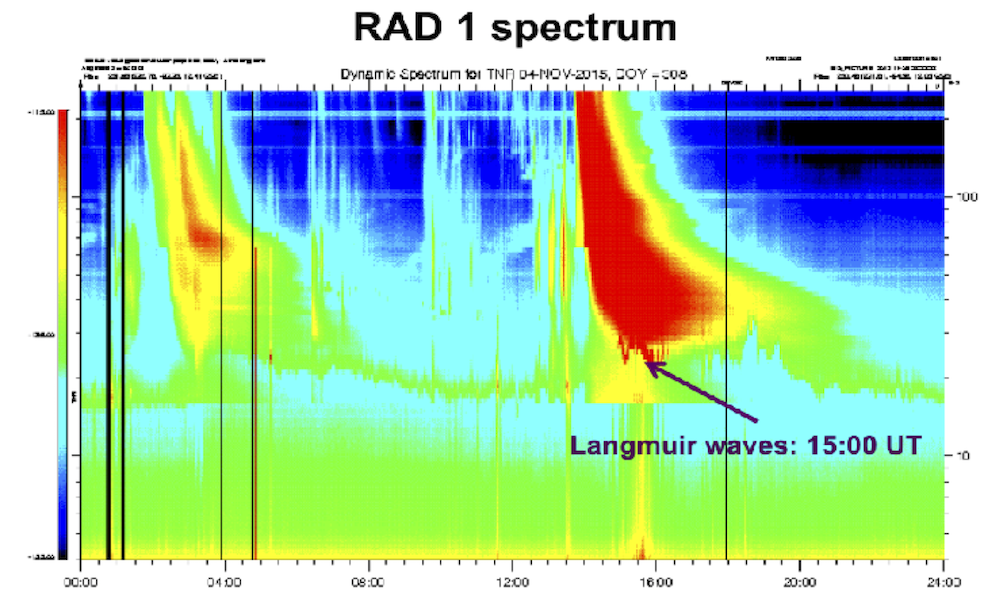} 
\caption{{\it Wind} WAVES/RAD1 dynamic spectrum  for 4 November 2015.} 
\label{Fig_Carolina_peaks}
\end{figure}

Next we analyse the {\it Wind} radio and plasma-wave data to constrain observationally the presence or absence of energetic electrons that reach $1$~AU near Earth. Figure \ref{Fig_Carolina_peaks} shows a partial RAD1 dynamic spectrum ($\approx 5 - 250$~kHz) for 4 November 2015, while Figure \ref{fig:radio_summary} shows the entire RAD1 (to ~1 MHz) and RAD2 ($\approx 1 - 14$~ MHz) domains. Figure \ref{Fig_Carolina_peaks} primarily shows some weak intermittent individual Type IIIs from 13:30 \,--\, 13:50~UT and then a saturated, longlasting emission that might be identified as a Type III-L burst \citep{cane_etal_2002}. However, Figure \ref{fig:radio_summary}a clearly shows multiple individual Type IIIs below $\approx 1$~MHz throughout the approximate period 13:30 \,--\, 14:30, although above about $\approx 4$~MHz separation into two groups appears reasonable. Evidence thus exists for Type III electron beams that are radio-quiet above $\approx 4$~MHz but radio-loud below $\approx 1$~MHz.

Figure \ref{Fig_Carolina_peaks} shows direct evidence for generation of Langmuir waves associated with the multiple interplanetary Type III bursts present. This evidence is the intensification of the ``plasma line'' near $f_{pe}$, corresponding to the green line near $20$~kHz from 08:00 to 14:00~UT that becomes red in the approximate period 15:00 $-$ 16:30~UT and ranges in frequency between about 20 and 30 kHz. Thus, during at least this latter period it appears that Type III electrons are present and unstable to the growth of Langmuir waves.

\section{Magnetic Field Configuration and Interplanetary Conditions}\label{plasma} 
\label{S6}

\subsection{Coronal Magnetic Fields and Heliospheric Connectivity}

\begin{figure}[ht] 
\centering
\includegraphics[width=1.0\columnwidth]{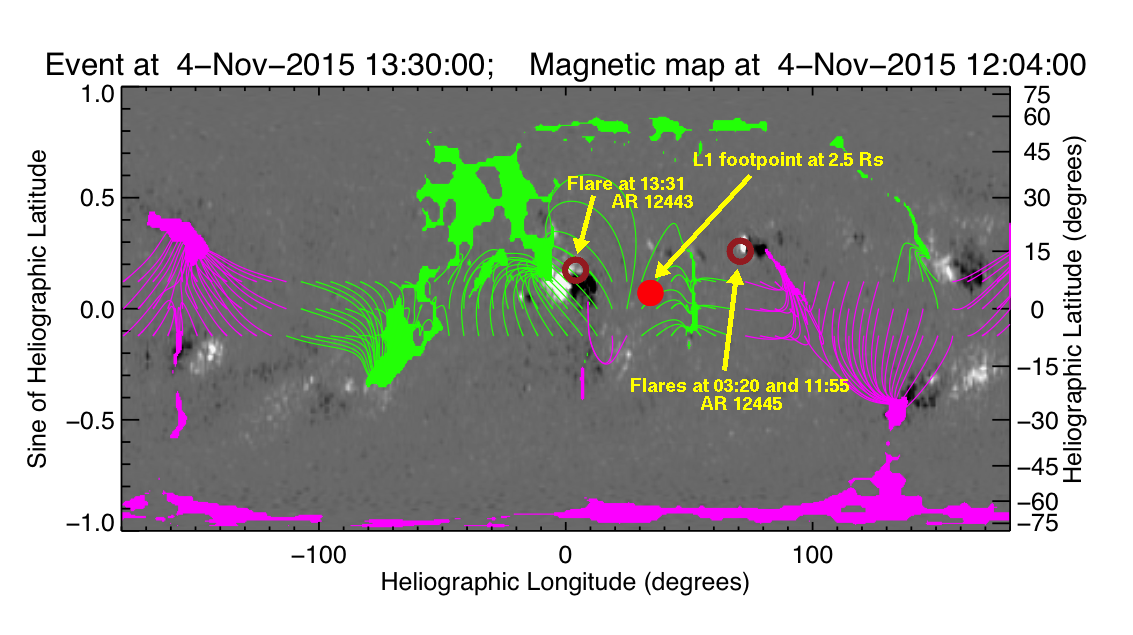} 
\caption{Predictions of a PFSS model with HMI photospheric magnetic field data for a synoptic magnetic map at the photosphere in Carrington coordinates around the time of the second flare, except that the Carrington longitudes are translated to Earth-view longitudes. The {\it open circles} show the photospheric locations of the flares with solar rotation corrected. The {\it red circle} is the footpoint of the magnetic field line traced from L1 to the source surface at $2.5 \rsun$ using Wind spacecraft velocity data and the Parker spiral (second method in the text).  Open field regions are marked in two colors ({\it green}: positive, and {\it pink}: negative).}
\label{nariaki_pfss_fig}
\end{figure}

Figure~\ref{nariaki_pfss_fig}, based on the PFSS model, as adopted by \citet{Schrijver:2003} and implemented in \textsf{SolarSoft}, predicts that the open field lines that reach the ecliptic plane at the source surface, placed at a heliocentric distance of $2.5 \rsun$, come from three places: the western periphery of AR~12445 (negative polarity -- purple lines), a coronal hole to the Southeast of the region (positive polarity -- green lines), and the eastern periphery of AR~12443 (positive polarity). Similar results were obtained (not shown) by the National Solar Observatory using its PFSS model with GONG magnetogram data.

Instead of using the PFSS model with photospheric magnetic field data, we can start at the Earth and use the solar wind speed observed by the {\it Wind} spacecraft and the Parker spiral model for ${\bf B}({\bf r})$ to estimate the heliographic coordinates on the source surface of the field line that crosses the Earth at a given time. From there, we can map the field line to the photosphere using the PFSS model. Using this approach, the field line that was connected to the Earth at 12:00 UT on 4 November was rooted in the coronal hole between AR~12443 and AR~12445, and its polarity matches that observed at L1. Using a more sophisticated heliospheric MHD simulation model (provided by Predictive Science, Inc.) shows that the footprint of the Earth-connected field lines became rooted in AR~12443 at 4 November 00:00~UT. This connection lasted until the next day. A note of caution is that the first eruption may distort the magnetic field configuration (and the other plasma structures) from the PFSS and Predictive Science, Inc. predictions for the second event.

\begin{figure}
\centering
\includegraphics[width=1.0\columnwidth]{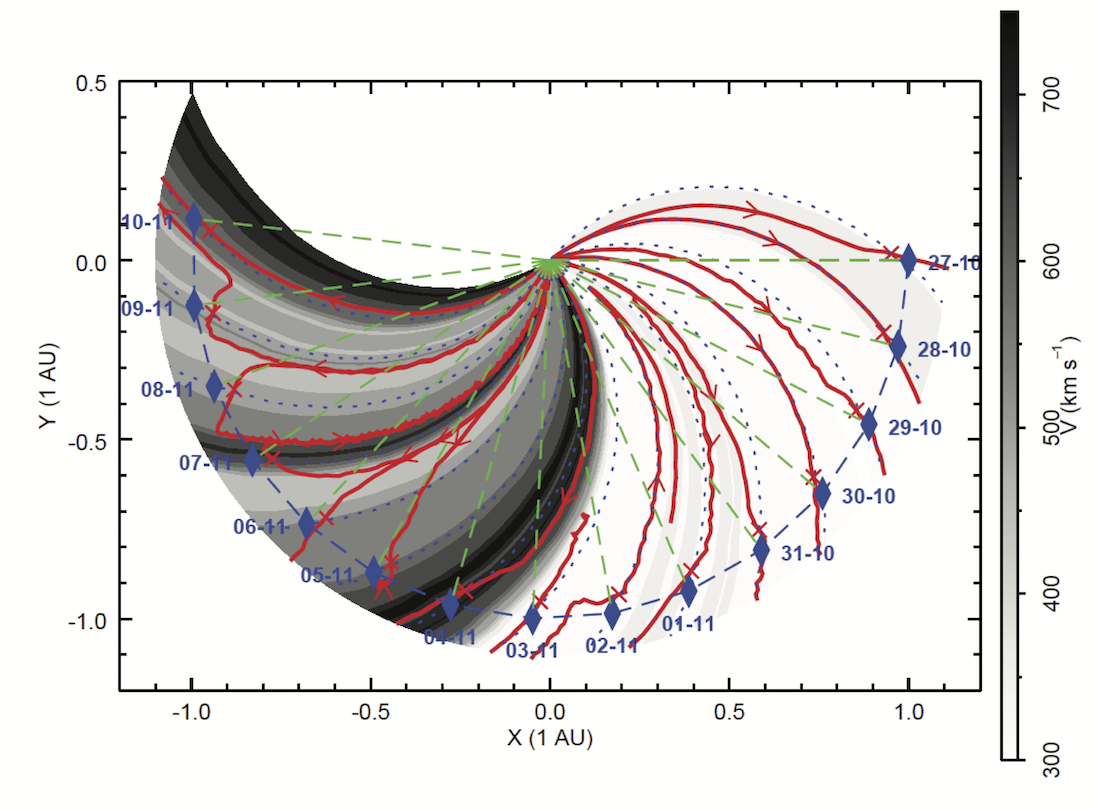}
\caption{Magnetic map in the solar equatorial plane for the interval 27 October to 11 November 2015, calculated using {\it Wind} spacecraft data and the approach of \citet{li_etal_2016}. {\it Dashed blue lines} show the nominal Parker spirals while {\it blue diamonds} show the position of the Earth on specific days. {\it Dashed green lines} show the Sun--Earth line for each day.}
\label{fig:BoLi_1}
\end{figure}

A different approach altogether is to extrapolate $1$~AU {\it in-situ} solar wind observations back in time and space. Figure \ref{fig:BoLi_1} maps the large-scale magnetic field lines and solar wind velocity streams in the solar equatorial plane, extending up to a distance of \mbox{2~AU} and constructed using a 2D solar wind model \citep{schulte_etal_2011, schulte_etal_2012} and the approach of \citet{li_etal_2016}. The process involves fitting {\it Wind} measurements of $B_\phi$, $B_r$, and $v_{r}$ at $1$~AU to an analytic model, permitting calculation of these quantities and the plasma density from $1$~AU to an inner boundary (nominally at the photosphere). The analytic model assumes the magnetic field to be frozen-in to the plasma, the wind sources to be constant over a solar rotation (so that the wind and its magnetic field lines form a constant pattern that rotates with the Sun, thereby not modelling CME effects properly), the flow speed to be constant along each streamline, and the plasma to corotate with the Sun at the inner boundary. The model and fitting procedure allow the magnetic field to be non-radial at the inner boundary.

On 4 November, Earth is near the bottom of Figure \ref{fig:BoLi_1}. The field line that reaches Earth on 4 November leaves the Sun about $45$\degree westward from the Sun\,--\,Earth line. It is quite closely Parker-like, albeit longer, but its neighbors are not: the nearest eastward field line is directed almost radially near $1$~AU (and so lies at an angle of order $45$\degree to the nominal Parker direction) and so initially moves westward rather than eastward, while the nearest westward field line does not proceed far Sunward due to it entering a region with magnetic field close to zero (note the opposite directions for the lines that reach $1$~AU between 2 and 4 November), and the next westward field line is part of a loop disconnected from the Sun that does reach 0.1~AU. These aspects suggest that electrons produced by Events 1 and 2 very close to the west-limb source (whether flare- or shock-produced) should not be magnetically connected to Earth on 4 November, unless there is substantial scattering or the acceleration region has a large angular width. Semi-quantitatively, it appears that scattering through $25 - 60 $\degree of longitude is needed, on considering the location of AR 12445 on the Sun and comparing the angular distances at $1$~AU between the corresponding field lines that connect to Earth's locations for 31 October to 2 November with the field line for 4 November. However, the disk center source for Event 3 should be magnetically connected to $1$~AU about $15$ degrees eastward of Earth (the angular distance between the field lines that connect to Earth on 4 and 5 November). Thus, Event 3 should be better magnetically connected than Events 1 and 2 but still not well connected.

\subsection{Interplanetary Plasma and Field Observations}\label{ipconditions}

\begin{figure}
\centering
\includegraphics[width=1.0\columnwidth]{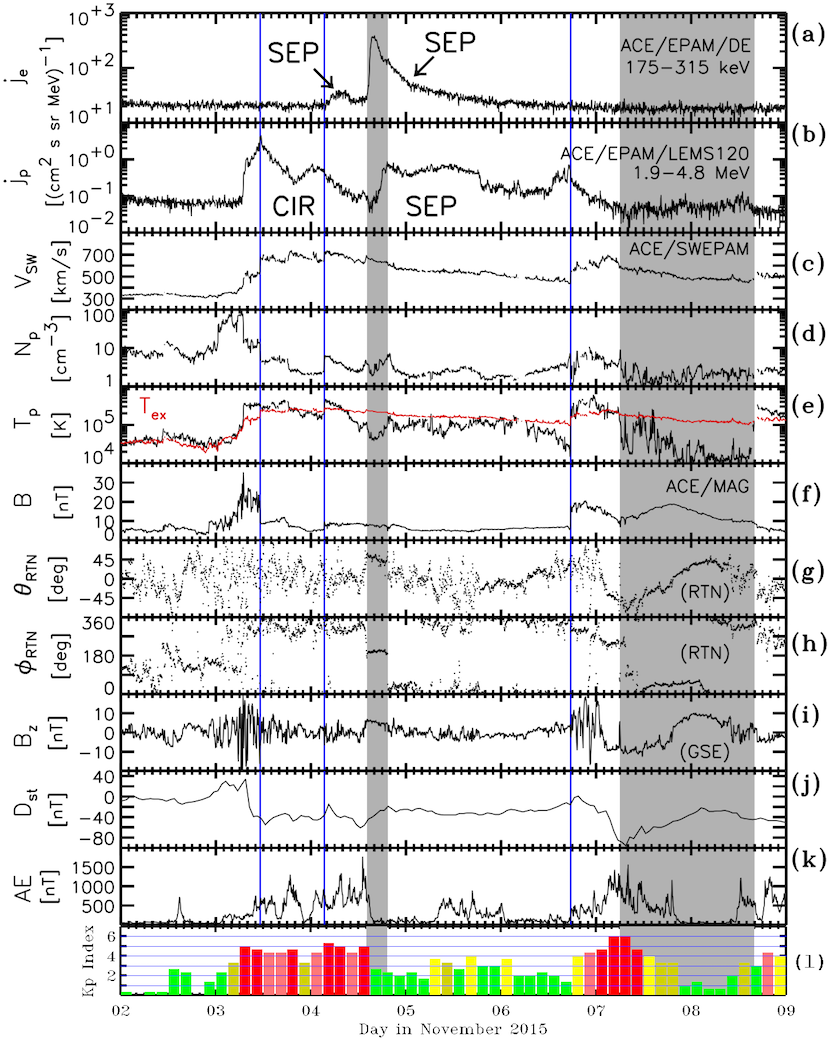}
\caption{Overview of in-situ observations: ({\bf a}) $175--315$~KeV electron intensities, ({\bf b}) $1.9--4.8$~MeV ion intensities, ({\bf c}) solar wind proton speed, ({\bf d}) solar wind proton density, ({\bf e}) solar wind proton temperature, ({\bf f}) magnetic field magnitude, ({\bf g}) magnetic field latitudinal angle, ({\bf h}) magnetic field azimuthal angle, ({\bf i}) $B_{z}$, the $z$ GSE component of the field, ({\bf j}) Dst index, ({\bf k}) AE index, and ({\bf l}) Kp index. {\it Blue vertical lines} identify shocks while {\it shaded areas} correspond to ICME periods. See text for details.}
\label{fig_context}
\end{figure}

Figure \ref{fig_context} overviews the in-situ plasma, magnetic field, and particle observations in the vicinity of the Earth from 2 November (DOY 306) to 9 November (DOY 313) 2015. The figure shows from top to bottom: (a) $175-315$~KeV electron intensities and (b) $1.9-4.8$ ion intensities observed by ACE/EPAM \citep{gold_etal_1998}, the solar-wind proton (c) speed, (d) density, and (e) temperature observed by ACE/SWEPAM \citep{mccomas_etal_1998}, the (f) magnetic-field magnitude and (g,h) field angles in Radial Tangential Normal (RTN) coordinates, and (i) $B_z$ component in the Geocentric Solar Ecliptic (GSE) frame observed by ACE/MAG \citep{smith_etal_1998}, and the (j) Dst, (k) AE, and (l) Kp geomagnetic activity indices. The red line in the temperature plot corresponds to the solar-wind proton temperature predicted for non-ICME periods using the solar wind speed and expression of \citet{Elliott_2012}. The vertical solid lines mark interplanetary shock passages observed in-situ and shaded regions indicate ICME periods identified by I.G. Richardson and H.V. Cane (\url{http://www.srl.caltech.edu/ACE/ASC/DATA/level3/icmetable2.html}).

The enhanced magnetic field and density region observed on 3 November corresponds to a stream-stream interaction region, produced when the high-speed stream (speeds $v_{sw} \approx 700$~kms$^{-1}$) observed during 3\,--\,4 November compressed the preceding slow solar wind ($v_{sw} \approx 320$~kms$^{-1}$). The first vertical line corresponds to a reverse shock associated with that CIR. This shock was coincident with the local peak of the energetic proton intensity increase associated with the CIR. 

A second interplanetary shock was observed on 4 November 03:24~UT, shortly before the onset of a weak solar electron event (see first label ``SEP'' in the top panel). An interval with ICME signatures was observed 10.8 hours after this shock, as indicated with the first shaded area from 14:10~UT to 19:26~UT in Figure \ref{fig_context}. These signatures included smooth magnetic-field rotation, low temperature and bi-directional solar-wind electron flux (the latter not shown). No significant increase of the magnetic-field magnitude was observed, suggesting that the spacecraft crossed close to one of the flanks rather than near the central part of the ICME. A second SEP event, showing electron and proton increases, was observed when the spacecraft was inside the ICME. The connection of these first two SEP events to Events 1\,--\,3 is discussed in detail in the next Section. Here we emphasise that the two shocks just discussed reach $1$~AU far too early 
be associated with the three solar events on 4 November 2015. 

We add for the purpose of the space weather discussion in Section \ref{S8} that a third interplanetary shock was observed on 6 November 17:35~UT, followed by an ICME covering the period 7 November 06:00~UT to 8 November 16:00~UT. This ICME showed very clear signatures, and it was likely the interplanetary counterpart of the CME ejected on 4 November in association with Event 3. The time elapsed is 52 hours and the averaged speed is $800$~km~s$^{-1}$. This average transit speed lies between the two plane-of-the-sky estimates of $\approx 600$ and $1200$~km~s$^{-1}$ estimated for the associated CME in Section 4.3. 

\section{SEPs}\label{seps} 
\label{S7}

\subsection{SEP Observations}

\begin{figure}[t]
\centering
\includegraphics[width=0.9\columnwidth]{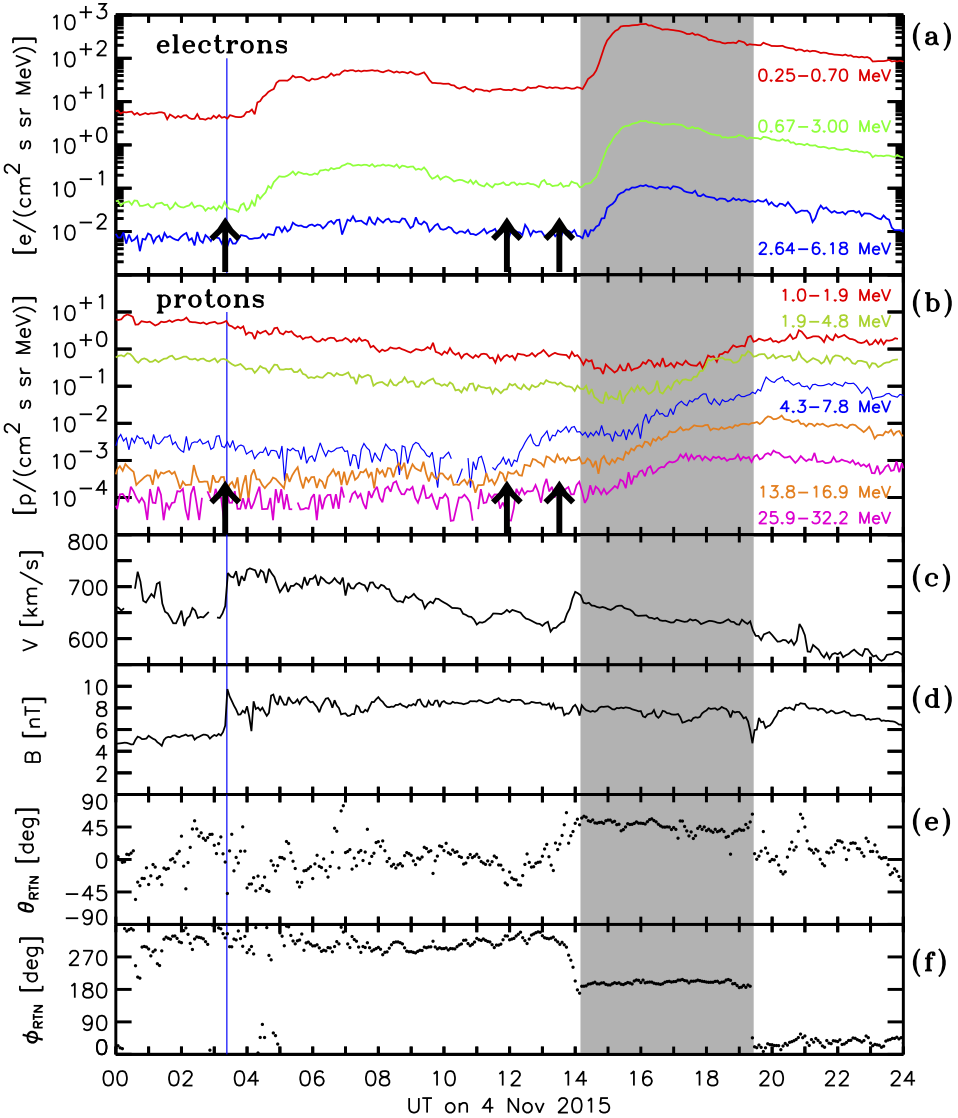}
\caption{Energetic electron and proton, solar-wind speed and magnetic-field vector observations during 4 November 2015, from instruments on SOHO and ACE at L1. {\it Thick vertical black arrows} mark the start times of Events 1-3.}
\label{fig_seps}
\end{figure}

Figure \ref{fig_seps} shows energetic particle, plasma flow, and magnetic field observations during 4 November 2015. From top to bottom the figure shows: (a) energetic electron intensities observed by SOHO/EPHIN \citep{muller-mellin_etal_1995} at three energy bands, (b) energetic proton intensities observed by ACE/EPAM, SOHO/EPHIN and SOHO/ERNE \citep{torsti_etal_1995} at five energy bands between $1$ and $32$~MeV, solar wind (c) speed, (d) magnetic field magnitude, and (e,f) magnetic field angular coordinates observed by ACE during 4 November. The shaded area corresponds to the first ICME shown previously in Figure \ref{fig_context}. The arrows in the first two panels mark the times of X-ray flares associated with the three events under study. During this period, the SOHO spacecraft was rotated 180 degrees from its nominal pointing, meaning that its particle instruments were pointing perpendicular to the nominal Parker spiral direction and missed the field-aligned particles expected to arrive first.

Two electron increases, with $0.25--0.74$~MeV onset times at 04:11~UT $\pm$2~minutes and 14:19~UT $\pm$2~minutes, were observed. The first increase, associated with Event 1, was clearly observed by the EPHIN instrument and the deflected electron channels of ACE (Figures \ref{fig_context} and  \ref{fig_seps}). The second electron increase, associated with Event 3, showed higher intensities than the first electron increase and was accompanied by energetic protons reaching energies up to 60~MeV. Both the electron events and the ion event showed clear velocity dispersion. 

The electron event 1 was not accompanied by an increase of the proton intensity, possibly because it was masked by the decaying proton intensity associated with the prior CIR. The small proton increase starting at 12:00 UT on 4 November, seen at the higher-MeV energies (see the blue curve in the second panel of Figure \ref{fig_seps}), started too early to be associated with Event 2. Therefore, there was no significant in-situ particle increase associated with Event 2. (Interestingly, however, modeling presented in Section 7.3 below suggests that this small proton event could be due to protons accelerated in Event 1 once its shock and CME are well away from the Sun.) During the period under analysis, there were no observations available from STEREO-B (communications with the spacecraft stopped in October 2014). STEREO-A was affected by reduced data return during the pass through solar conjunction, but the beacon data available show that the two SEP increases observed at L1 and described in Figures \ref{fig_context} and  \ref{fig_seps} were not observed by STEREO-A (at that time located at a heliographic longitude of $\approx 168$ degrees). Thus the observed SEP electron events do not cover $360$\degree of heliolongitude at 1~AU.  

\begin{figure}[t]
\centering
\includegraphics[width=0.9\columnwidth]{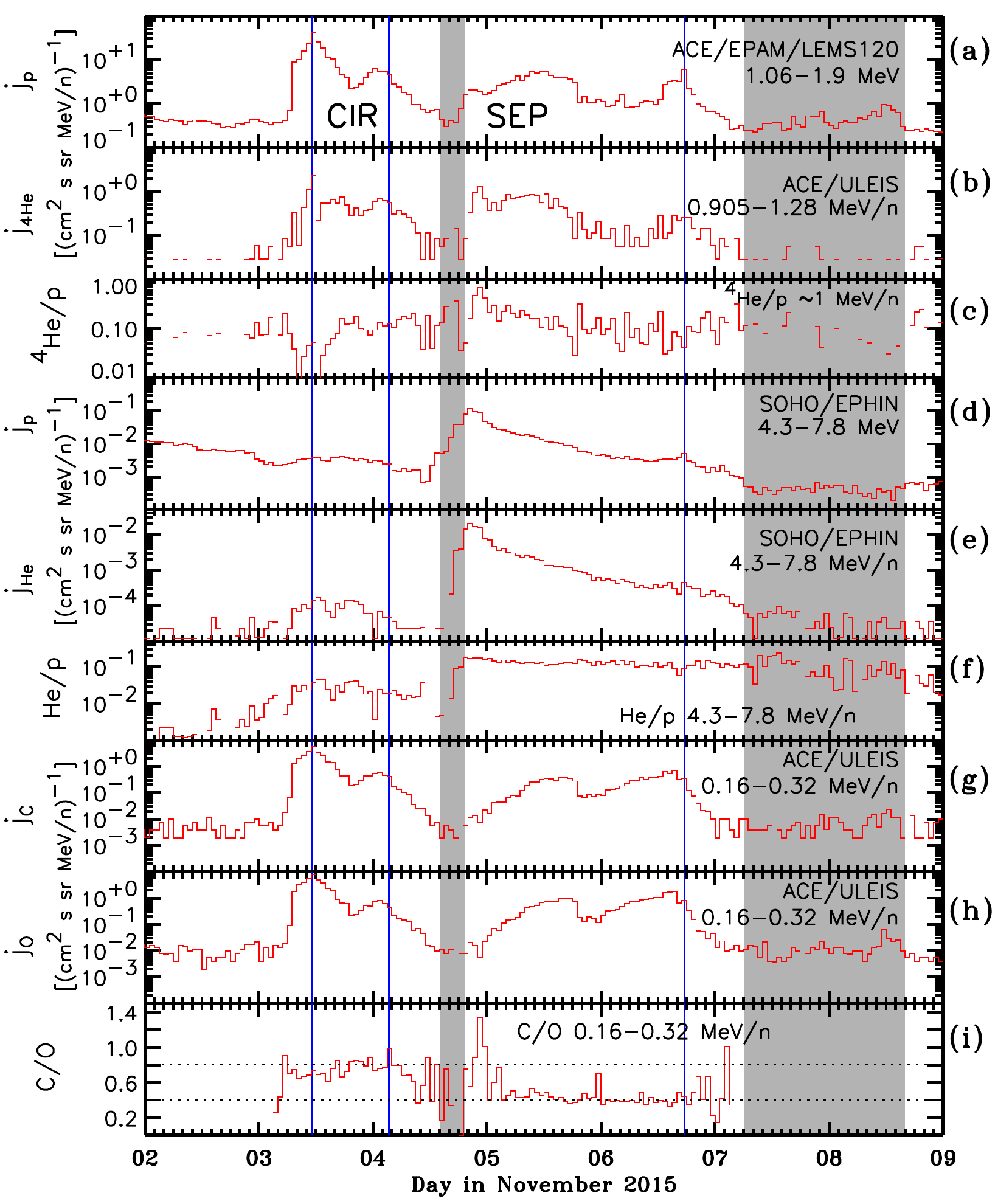}
\caption{Intensities and abundance ratios of multiple energetic ions observed by ACE/EPAM, ACE/ULEIS and SOHO/EPHIN between November 2 and November 9, 2015: {\bf a}) $1.06-1.9$~MeV proton intensity, {\bf b}) $0.905-1.28$~MeV/n helium intensity, {\bf c}) helium to proton abundance ratio at  $~1$~MeV/n, {\bf d}) $4.3-7.8$~MeV proton intensity, {\bf e}) $4.3-7.8$~MeV/n helium intensity, {\bf f}) $4.3-7.8$~helium to proton ratio, {\bf g}) $0.16-0.32$~MeV/n carbon intensity, {\bf h}) $0.16-0.32$~MeV/n oxygen intensity, and {\bf i}) $0.16-0.32$~MeV/n carbon to oxygen ratio.  Increases associated with CIRs and SEPs are labeled, as are ICME periods ({\it shaded regions}) and shocks ({\it blue vertical lines}).}
\label{fig_seps_comp}
\end{figure}

Figure \ref{fig_seps_comp} shows the intensities and abundance ratios of multiple energetic ions observed by ACE/EPAM, ACE/ULEIS and SOHO/EPHIN between November 2 and November 9 in 2015. Specifically the figure provides data for protons, helium (technically both $^{4}$He and $^{3}$He, although usually the fraction of $^{3}$He is expected to be negligible), carbon, and oxygen. The shaded areas correspond to ICME periods. While hypothetical SEPs ions from Event 1 might be masked by the CIR-associated increase starting on November 3, all of these species showed clear increases associated with Event 3. The He/H ratio remained close to 0.1 during the whole of Event 3, which corresponds to the typical values found during impulsive (flare-associated) SEP events (see,  {\it e.g.}, Reames 1999). The C/O flux ratio clearly separates the periods with CIR- and SEP-related energetic particles, as found previously \citep{mason_sanderson_1999}, providing additional arguments against Events 1 and 2 producing significant SEP ions. Note that the composition signatures of gradual and impulsive events can sometimes be blurred \citep[{\it e.g.}]{cohen_2006}), as apparently found here.

\subsection{Electron Anisotropy Observations and Modeling of SEPs related to Event 3}
\label{sect_electron_SEPs}

\begin{figure}[t]
\centering
\includegraphics[width=0.6\columnwidth]{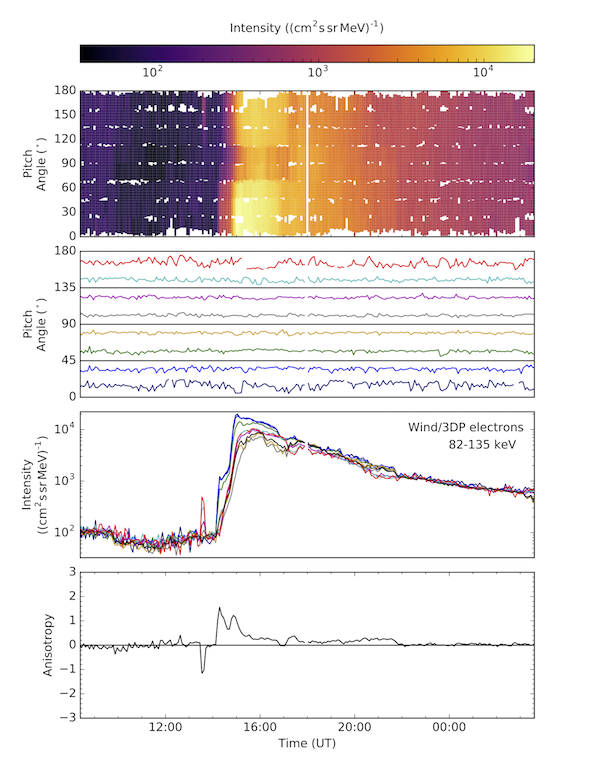}
\caption{Energetic electron observations by {\it Wind}/3DP. From top to bottom: Pitch-angle distributions with color-coded intensity, pitch-angle corresponding to each pitch-angle bin, electron intensities observed in each bin, and the first-order anisotropy index. (The peaks near 13:30 $--$ 13:45~UT in the bottom two panels are likely solar-flare light contamination and should be ignored.)}
\label{fig_anisotropy1}
\end{figure}

\begin{figure}[t]
\centering
\includegraphics[width=0.6\columnwidth]{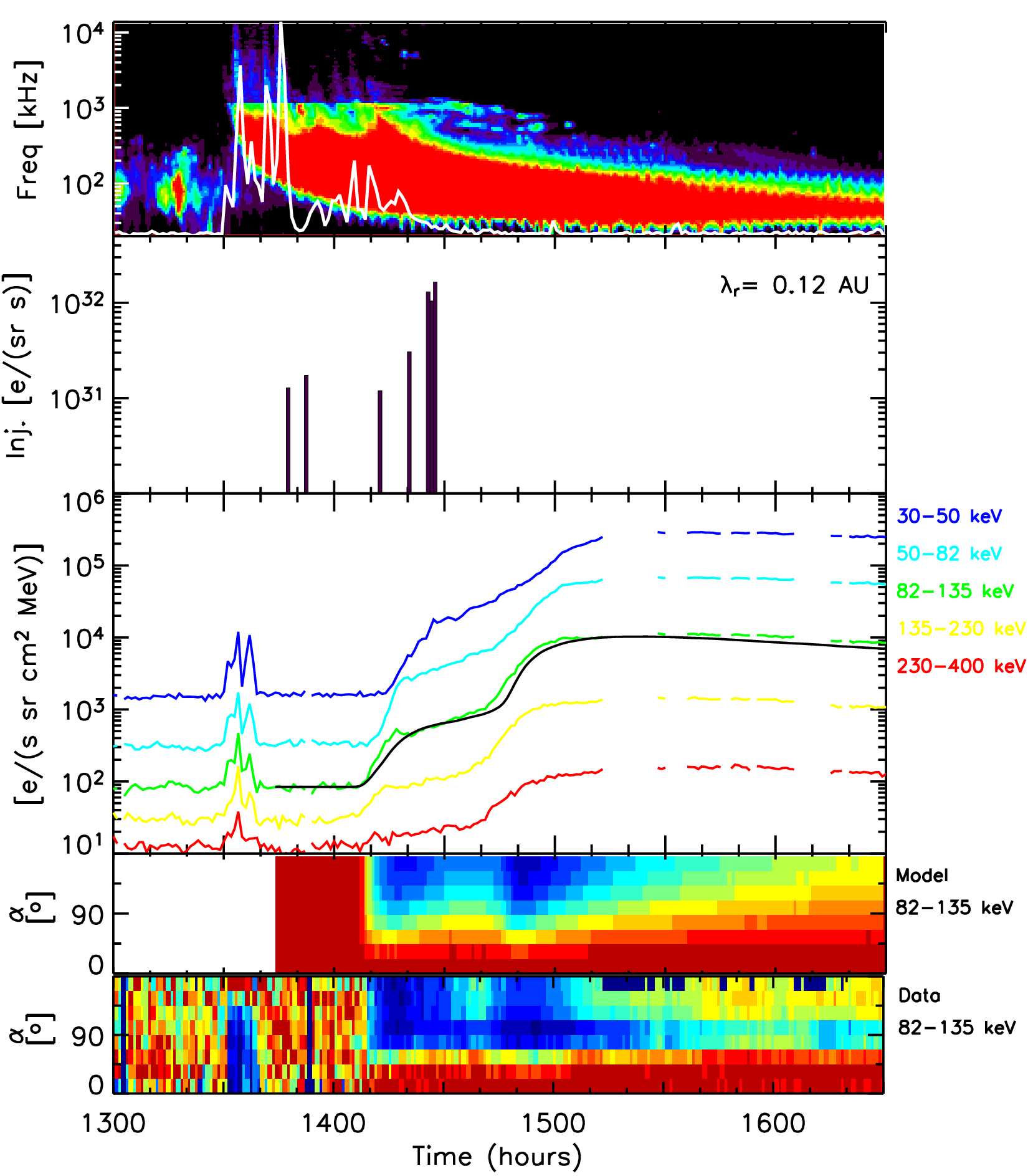}
\caption{{\it From top to bottom}: Radio spectra observed by {\it Wind}/WAVES (colour-coded by intensity, with red large and black background) and 13 MHz microwave emission (white curve, relative intensity) on 4 November 2015 (for comparison purposes with the particle data, the emissions were shifted by $-500$~s);  $82 -- 135$~KeV electron source profile deduced at $2 \rsun$; omni-directional intensities observed by {\it Wind}/3DP in the energy range $30 -- 400$~KeV, with the black curve showing the best fit to the $82 -- 135$~KeV data for $\lambda_r = 0.12$~AU; the two bottom panels show the modeled and observed $82 - 135$~KeV electron pitch-angle distributions normalized to the maximum value ($1 =$ {\it red} to $0 =$ {\it blue}).}
\label{fit_electrons}
\end{figure}

The electron data above $250$~KeV in Figure \ref{fig_seps} show only a single onset for Event 3. However, at the lower energies presented in Figures \ref{fig_anisotropy1} and \ref{fit_electrons} the electron event has a first peak approximately ten minutes after the onset, followed by a flat interval and then a second major increase of the particle intensity. Furthermore, analyses presented next of {\it Wind}/3DP sectored data show two clear episodes of velocity dispersion during the rising phase of the two peaks. ({\it SOHO}'s orientation prevents similar analyses for this period.) This two-step electron rise is unlikely to be due to local effects because it was observed by both the {\it Wind} and {\it ACE}  spacecraft, at that time separated by $130$~Earth radii and both immersed in the ICME and observing a steady field orientation. 

Figure \ref{fig_anisotropy1} shows the pitch-angle distributions of $82--135$~KeV electrons observed by {\it Wind}/3DP in association with Event 3. During the rising phase of the two peaks, {\it Wind}/3DP observed pitch-angle distributions peaking at small pitch-angles, indicating a good magnetic connection to the solar source region. Moreover, the first order anisotropy index shows a double-peak shape in agreement with the two-step increase of the intensities. These observations suggest that the energetic electron event was composed of two successive groups of injections separated by $\approx 30$ minutes. Later on, after about 15:00~UT, the non-monotonic evolution of the pitch-angle distributions ({\it i.e.} the observation of bidirectional distributions with an increase of sunward propagating electrons) signals that the spacecraft was inside an ICME. These electrons may have been reflected by the converging magnetic-field lines at the opposite ICME leg or by a reflecting magnetic barrier located significantly beyond 1~AU. In summary, the activity on 4 November 2015 was accompanied by an electron event (Event 1) that occurred in the sheath region between an ICME shock and its driver, and by an SEP event with enhanced proton and electron intensities (Event 3) that occurred within the magnetic obstacle of the ICME.

We modeled the early phase of the $82--135$~KeV electron event observed by {\it Wind}/3DP using simulations of the interplanetary transport of solar energetic particles, followed by the optimization of the injection and transport parameters. The transport model \citep{AguedaEtAl08} solves the focused transport equation \citep[see][for the full equation]{Ruffolo95}, which is essentially one-dimensional along a magnetic field line. It assumes the electron solar source at $2 \rsun$ and an Archimedean spiral magnetic-flux tube connecting the Sun and the spacecraft defined by the solar wind speed measured {\it in-situ}. (This approximation is not unreasonable for standard particle transport from the Sun but ignores the electrons observed moving Sunwards and this period's ICME environment.) Interplanetary pitch-angle scattering is parametrized assuming a pitch-angle diffusion coefficient that resembles the predictions of the ``standard model'' \citep{Jokipii66,JaekelEtS92}: the mean free path characterizes the degree of pitch-angle scattering. Following previous works \citep[{\it e.g.}][]{KallenrodeEtAl92}, we take the electron radial mean free path, $\lambda_r$, to be spatially constant.
We used the {\sf SEPinversion} software available in SEPServer (\url{http://server.sepserver.eu}) to infer the release time history and the value of the electron radial mean free path. {\sf SEPinversion} uses an inversion approach to fit the observations and it allows an estimation of the timing and intensity of the release without any {\it a priori} assumption on the profile.

Figure \ref{fit_electrons} shows the electromagnetic and particle data together with the best possible fit inferred using {\sf SEPinversion}. The best fit (second panel) is obtained assuming $\lambda_r = 0.12$~AU and multiple electron injections that occur in two primary groups; this is sometimes loosely called a two-episode release time profile, but notice that the best-fit model contains at least six electron injections, not two. The finding of two primary groups of releases is qualitatively consistent with the two-episode injection scenario signalled by the anisotropy index (Figure \ref{fig_anisotropy1}, bottom panel). 
 
Figure \ref{fit_electrons} shows that the model results cannot quantitatively explain the injection timing, as the two groups of inferred injections start $\approx 10 -- 20$ minutes after the beginning of the radio emission. The exact timing of the electron releases with respect to the electromagnetic emissions depends in part on the length of the interplanetary path, with a longer path allowing the injections to move earlier in time and so agree better with the radio events.  The analysis in Figure \ref{fit_electrons} assumes a nominal Parker spiral, whereas the field line length may be different. For instance, since the electrons were detected within an ICME, the field lines may be longer because of twisting \citep[but see]{kahler_etal_2011}. Alternatively, the field line that reaches Earth on 4 November in Figure \ref{fig:BoLi_1} is predicted to have a length of $1.2$~AU, while the nominal spiral is $1.1$~AU long. This would allow the electrons to be released $\approx 0.1/1.1\times 25 =   2.5$~minutes earlier, corresponding to moving the releases in Figure \ref{fit_electrons} to the left by $2.5$~minutes. Since the impulsive coronal radio emissions start $\approx 10 - 20$~minutes earlier than the first set of predicted electron releases, the increased field line lengths in Figure \ref{fig:BoLi_1} are insufficient to explain the timing difference. Instead, an enhanced field-line length in the range $1.5 -- 1.9$~AU is required, a clear quantitative problem. 

The two groups of releases inferred from the modeling at first glance appear to agree with the two main groups of radio emissions apparent in the $13$~MHz emissions in Figure \ref{fit_electrons} (top panel, white curve) near $\approx$ 13:30 \,--\, 13:45 and 13:50 \,--\, 14:20~UT. However, Figure  \ref{fig:radio_summary} clearly shows multiple individual Type IIIs below $\approx 1$~MHz throughout the approximate period 13:30 \,--\, 14:30~UT, although above about $\approx 4$~MHz separation into two groups appears reasonable. Evidence thus exists for Type III electron beams that are radio-quiet above $\approx 4$~MHz but radio-loud below $\approx 1$~MHz, as modelled by \citet{li_and_cairns_2012,li_and_cairns_2013}. The radio emissions and the Langmuir waves and energetic electrons observed at $1$~AU show release of electrons into interplanetary space that are connected to L1. The concept of Type III cannibalism \citep{li_etal_2002} is a plausible way to explain qualitatively why the multiple distinct decametric Type IIIs observed correspond to only two distinct signatures in the electron pitch-angle data.


\subsection{SEP Proton Event Modeling}

As summarised in Section 2, there are many challenges involved in modelling specific SEP events quantitatively, with many formalisms, models, and approximations involved. The approach adopted here is to present a modelling challenge to the community for the events of 4 November 2015 and to present one unusual analysis (that includes drift effects but assumes a constant, energy-independent, mean free path along Parker spiral fields) whose surprisingly good results are intended to stimulate the community into a concerted effort.

Analysis and interpretation of energetic protons observed from the 4  November 2015 events were hampered by the rotation of SOHO, which resulted in the ERNE instrument pointing perpendicular to the mean Parker spiral direction. Thus, applicable flux anisotropy profiles are not available and an analogue to the inversion method used for electrons in Section \ref{sect_electron_SEPs} is not possible. Similarly, although {\it Wind}/3DP observed flux anisotropies below \mbox{11 MeV} (not shown here), the observations were inadequate for producing reliable inversion results.

An alternative to applying an inversion method to data to obtain the particle transport parameters and mean free paths (usually assuming transport confined to a single field line) is to allow for drifts and cross-field motion and to assume values for the prevalent mean free paths. This approach can also place constraints on the magnetic connectivity between the SEP source and observing location. Accordingly, we performed a number of simulations of SEP 
(proton) transport for the 4 November 2015 events using the full-orbit propagation model of \citet{Marsh2013}, capable of accounting for drifts and deceleration effects. The input database was constructed using the approximation of a constant, energy-independent, proton mean free path of $0.3$~AU.

The results of the simulations were fed into the SEP forecasting tool described by \citet{Marsh2015}, assuming a solar wind speed of $700$~km s$^{-1}$ (see Figure \ref{fig_context}) and the associated Parker-spiral magnetic field connectivity for an assumed unipolar outward-pointing field at the coronal base. (See the comments in Section~7.2 about the actual magnetic environment.) The tool generates synthetic time profiles for protons at $1$~AU by assuming injection at a flare-related shock, with a width of $48$ degrees centered at the flare location, and with the injection constrained by the observed magnitude of the soft X-ray flare and a published correlation between soft X-ray flare magnitude and peak SEP intensity \citep{Dierckxsens2015}. 

\begin{figure}[h]
\centering
\includegraphics[width=0.8\columnwidth]{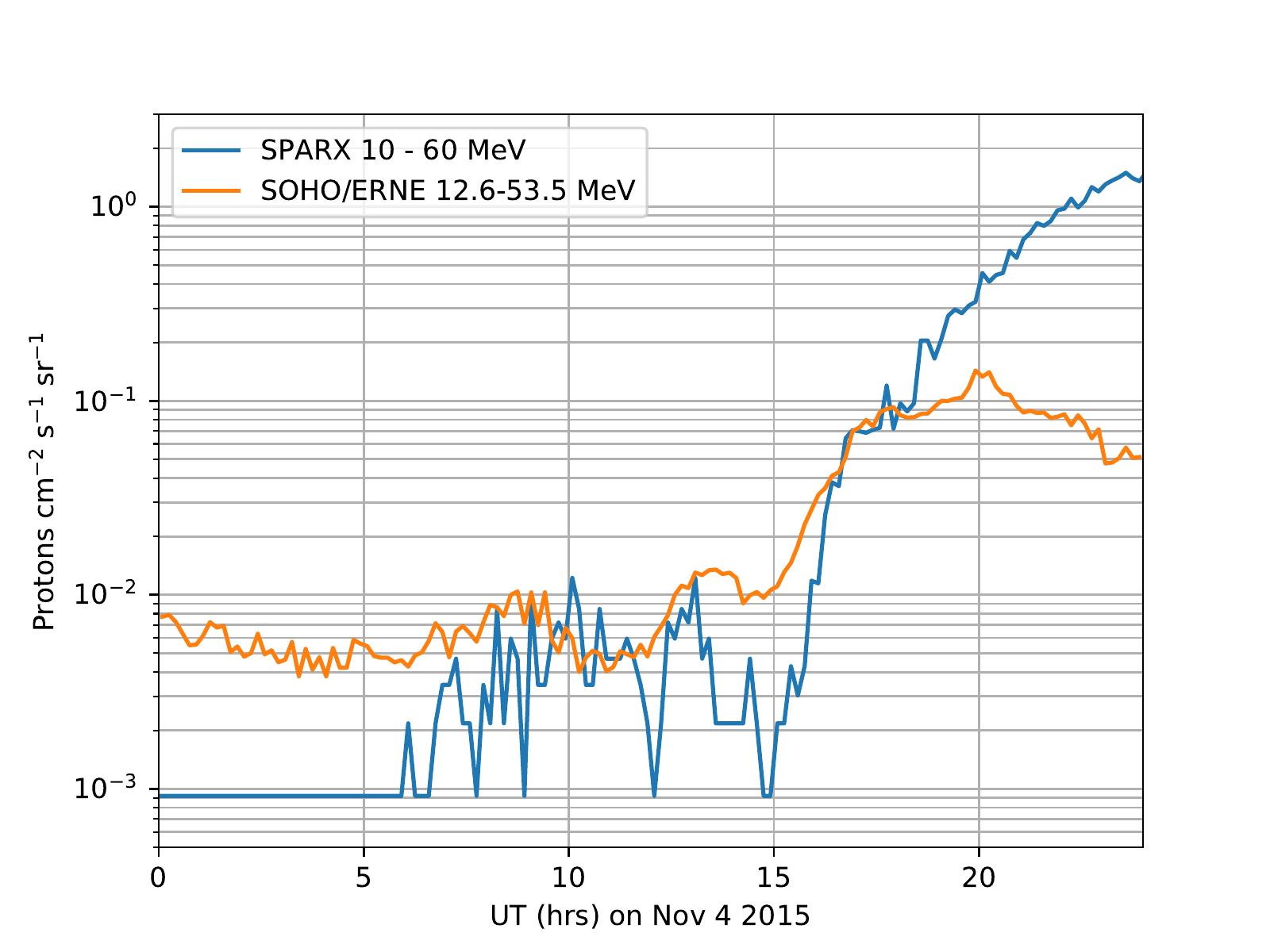}
\caption{Combined integral ERNE HED proton observations on 4 November for \mbox{12.6 -- 53.5~MeV} ({\it orange}) are compared with the temporal profile predicted for \mbox{10 -- 60~MeV} protons by the SPARX forecasting software ({\it blue}), assuming a mean solar wind speed of \mbox{700~km s$^{-1}$}. At the time of this event, SOHO was rotated 180 degrees, so ERNE was pointing perpendicular to the mean magnetic field.}
\label{fig:SPARX_ERNE}
\end{figure}

For the foregoing parameters, the simulation tool predicts only barely observable (proton) SEP levels for Events 1 and 2, as shown by the fluxes predicted before ~15:00 UT in Figure~\ref{fig:SPARX_ERNE}, not inconsistent with the observations in Figure \ref{fig_seps}. The situation is different for Event 3, for which Figure \ref{fig:SPARX_ERNE} displays both the flux of \mbox{12.6 -- 53.5 MeV} protons observed at SOHO/ERNE and the flux of \mbox{10 -- 60 MeV} protons predicted by the simulation tool. (For this event an additional static background flux of \mbox{$10^{-3}$ protons cm$^{-2}$ s$^{-1}$ sr$^{-1}$} was added to the tool's predictions to produce the displayed results.)  

Figure \ref{fig:SPARX_ERNE} suggests that the \citet{Marsh2015} model predicts the existence, timing, and qualitative size (within one to two orders of magnitude) of Event 3's SEPs quite well. Additional support for the model working surprisingly well (despite ignoring the ICME environment and non-Parker field lines) is that running the simulation for $v_{sw} = 500$~km~s$^{-1}$, corresponding to more common solar-wind speeds, and associated Parker magnetic connectivity leads to the prediction of SEPs for Events 1 and 2 but no SEPs for Event 3. These aspects are both inconsistent with the observations in Figures \ref{fig_seps} and \ref{fig:SPARX_ERNE}. 
    
More detailed comparisons of the predicted and observed proton fluxes in Figure \ref{fig:SPARX_ERNE} illustrate the important roles of magnetic connectivity.  First, while the observed and predicted onsets are very similar (near \mbox{15:30 UT}), the peaks are not. The decline after the observed peak at approximately \mbox{20:00 UT} coincides with a return to outwards magnetic polarity (near 19:45 UT on 4 November in Figure \ref{fig_context} that is not included in the model. Second, the time profiles for Event 3 and the low observed and predicted proton fluxes for Events 1 and 2 shown in Figure \ref{fig:SPARX_ERNE} are strong evidence that the proton SEP event observed at $1$~AU on 4 November is due to Event 3, located near the central meridian and that the effective solar-wind speed $v_{sw} \approx$ 700~km~s$^{-1}$. Third, for Event 3, connectivity to Earth (SOHO) is sub-optimal, with the early phase of the event not resulting in any flux. Rather, connected field lines sweep over the observer at \mbox{1~AU} after the flare has occurred, consistent with a near-isotropic proton population impacting the SOHO/ERNE detector (which was pointing perpendicular to the mean magnetic-field direction). Differences in onset time between simulations and observations may be due to variations in the exact shape of the interplanetary magnetic field, the proton mean free path, or the spatial extent of the acceleration region at the Sun. Fourth, we further deduce that the decay of the fluxes may be due to the non-trivial magnetic connectivity, and the transition between magnetic-field polarities. This is consistent with recent studies in SEP propagation, which confirm that magnetic polarity reversal boundaries are efficient at preventing particles from crossing them \citep{Battarbee2017}.

Further work might follow several approaches: First, to repeat the foregoing analysis for multiple mean free paths so as to find the optimum value, requiring extensive computational resources not available to our collaboration. Second, to include the non-Parker-like fields in Figure \ref{fig:BoLi_1} into the foregoing orbit calculations \citep{zhang_etal_2009,Marsh2013} and to compare the theoretical results, thereby directly assessing the importance of non-Parker fields for this event. Third, to perform the more standard analyses that include diffusive shock acceleration, scattering, magnetic focusing, and cross-field transport in one dimension \citep{li_etal_2009,verkhoglyadova_etal_2010, vainio_etal_2014, droge_etal_2014, he_wan_2015, hu_etal_2017} and in two dimestions \citep{hu_etal_2017}, thereby allowing assessment of the different physics included and of multi-dimensional effects. Finally, the CMEs, shocks, and background medium  should be modelled as accurately as possible using multi-dimensional MHD simulations \citep{kozarev_etal_2013,schmidt_etal_2013,schmidt_etal_2016,hu_etal_2017} and coupled with particle acceleration formalisms \citep{verkhoglyadova_etal_2010,kozarev_etal_2013,hu_etal_2017} to predict the SEP properties, with comparisons elucidating the roles of the shock evolution, 3D background plasma, non-Parker magnetic fields, and the physical processes considered.

\section{Space Weather: Event Timing and Characteristics}
\label{S8}

Event 3 occurred near the centre of the solar disk and was associated with a broad CME and extended dimming regions most clearly seen in AIA images in the $193 \AA$ and $211 \AA$ channels (Figure \ref{fig:eruption_summary}). Its angular width was estimated to be $226$\degree by CACTus \citep{Robbrecht04}. Such a CME is generally thought to be Earth-directed. Accordingly, NOAA/SWPC made predictions of the arrival of the CME-driven shock wave at the Earth. This was part of their routine solar wind prediction using the WSA-ENLIL model (see www.swpc.noaa.gov/products/wsa-enlil-solar-wind-prediction). A pressure pulse representing the CME is inserted at ENLIL's inner boundary at $21.5 \rsun$. The required CME parameters consist of the time at which the CME passes the heliocentric distance of $21.5 \rsun$, the direction (latitude and longitude), width (half angle), and radial velocity of the CME.  After Event 3, three runs were made with different CME parameters as we can see in the NGDC archive (www.ngdc.noaa.gov/enlil).  The parameters used in these three ENLIL runs are summarized in Table \ref{table_nariaki}.

\begin{figure}
\centerline{\includegraphics[width=0.9\columnwidth]{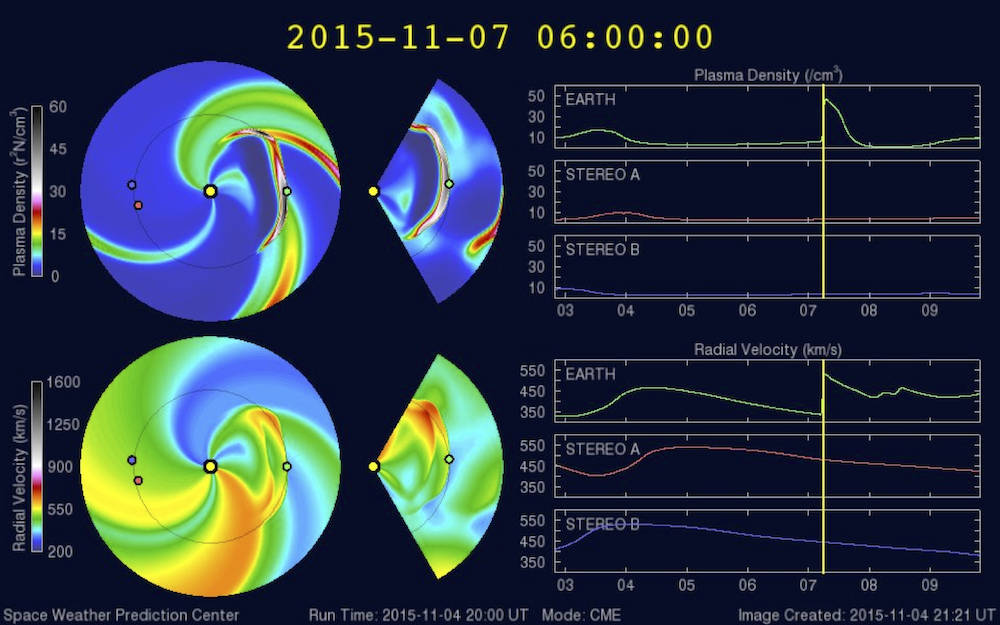}}
\caption{Snapshots of spatial maps in ({\it left}) the ecliptic plane and ({\it middle}) the meridional plane through the Sun-Earth line, as well as ({\it right} temporal variations at Earth's location for the ({\it upper row}) plasma density and ({\it lower row}) plasma speed. According to this run, the predicted shock-arrival time was 06:00~UT on 7 November 2015.}
\label{figure_enlil}
\end{figure}

The prediction that was closest to the actual shock arrival (6 November 2015, 17:34~UT) was made at 20:00~UT on 4 November ($\approx$6.5 hours after the flare onset).  This is shown in Figure \ref{figure_enlil}. The predicted arrival was 06:00~UT on 7 November.  The other two runs predicted later arrivals (15:00~UT on 7 November and 03:00~UT the next day), indicating the present capability for predicting the CME shock arrival time is $\approx \pm 12$ hours \citep{vrsnak_etal_2014}. Note that the CME parameters were obtained using only Earth-based observations, since STEREO A and B had not resumed operations after their solar-opposition passes earlier in 2015. Therefore, there were large uncertainties in obtaining the parameters of the cone model. Another source that affects the CME propagation in heliospheric simulations is how well the ambient solar wind is characterized \citep{vrsnak_etal_2013}. Including the CMEs associated with Events 1 and 2, there were several CMEs of different sizes and speeds within a few days before Event 3, so it is possible that the actual heliosphere encountered by Event 3's CME was significantly different than modelled in the ENLIL simulations. 

A different method to estimate interplanetary travel times of ICMEs was proposed by \citet{CSM:Kle-15} and \citet{CSM:al-17}. These authors used the fluence of the soft X-ray or microwave (8.8~GHz) burst associated with the CME to estimate the propagation speed of the CME front. This speed estimate is fed into the empirical propagation model of \citet{Gop:al-01b} to predict the arrival time at $1$~AU of the ICME's magnetic cloud, rather than the shock. The SXR burst of 4 November 2015 in the $0.1 - 0.8$~nm band rose to a maximum of about $3 \cdot 10^{-5}$~W~m$^{-2}$ within about 22~minutes, producing a fluence of about $2.6 \cdot 10^{-2}$~J~m$^{-2}$. The CME speed inferred from the empirical relationship given by Equation~2 of \citet{CSM:Kle-15} is 950~km~s$^{-1}$. The estimated arrival time at $1$~AU is then 05:20~UT on Nov 07.

ACE/SWEPAM solar wind data (Figure \ref{fig_context}) show that a time interval where the proton temperature is less than half the expected value for a standard solar wind stream \citep{ell05} starts at 6~UT (between 5:53 and 6:21~UT) on 7~November. At about the same time the solar wind speed starts a systematic decrease which lasts over several hours. These indications of the arrival of the ICME at ACE are close to the time predicted using the SXR burst fluence, within $\approx 1$~hour.  

Microwave observations from the Sagamore Hill station of the {\it RSTN} show an impulsive burst ($\approx$ 13:37\,--\,13:47~UT) that rises to a peak flux of $782$~sfu at $8.8$~GHz and has a weak longer-lasting tail (not shown). The weak tail is probably thermal bremsstrahlung. The impulsive burst has a fluence of $1.5 \cdot 10^{-17}$~J~m$^{-2}$~Hz$^{-1}$. Equation~1 and the coefficients for $9$~GHz in Table~2 of \citet{CSM:al-17} translate this into a CME speed of  $460$~km~s$^{-1}$, which leads to an arrival prediction of 16:35~UT on 8~November. This is clearly much later than observed: the microwave method leads to a considerable underestimation of the CME speed in the corona. Comparing the predictions using the SXR burst fluence {\it versus} the microwaves for Event 3 suggests that the SXR method is considerably more accurate in this particular case. This agreement is exciting and suggests that the methods of \citet{CSM:al-17} need to be tested in detail for additional events.

Once at Earth the CME related to Event $3$ produced a set of space weather phenomena. Indeed, Figure \ref{fig_context} shows multiple southwards excursions of $B_{z}$ in the sheath region between the shock and the tangential discontinuity in front of the CME material, together with a smooth rotation of $\theta_{RTN}$ and $B_{z}$ changing from southwards to northwards. A standard geomagnetic substorm occurred with a sudden storm commencement and a significant intensification of the ring current and auroral zone activity. Specifically, the minimum $D_{st}$ was $\approx -100$~nT and the AE and Kp indices were enhanced from the shock arrival until about 2000~UT on 7 November (maximum values $\approx 1500$~nT and $6$, respectively) when $B_{z}$ returned to being northwards.
 
A major space-weather effect of Event 3 was that the Swedish aviation radar systems were severely impacted by an extremely strong radio burst at GHz frequencies \citep{Opgenoorth:2016}. The effect was also seen in other European countries. The extreme intensity of the radio burst was only in a limited frequency range. A detailed study by \citet{marque_etal_2018} shows that the perturbations coincide in time with two intense peaks of the radio bursts near the operational frequencies of the secondary air traffic control radar. This makes the radio burst the prime candidate to explain the Swedish air traffic incident. Such radio bursts may be outside the usual space-weather forecasts. At least, the impact of such events appears to be both temporally and spatially limited.

\begin{table}
\caption{CME parameters used for ENLIL runs by NOAA/SWPC}
\begin{tabular}{ccccccc}     
\hline
ID$^{a}$  & $t_{CME}$$^{b}$  & Lat. &
Long. & Half Angle & Vel.$^{c}$ & Time (predicted)\\
1 & 4 Nov 19:00 & 11 & 7 & 53\degree & 828 & 7 Nov 06:00\\ 
2 & 4 Nov 20:29 & -12 & 35 & 38\degree& 584 & 8 Nov 03:00\\ 
3 & 4 Nov 19:28 & 2 & 12 & 48\degree & 762 & 7 Nov 15:00\\ 
\hline
\end{tabular}
\noindent

a. The ID corresponds to the time at which ENLIL was run: 1. 20:00 on 4 November,\\
\hspace{0.1cm} 2. 00:00 on 5 November and 3. 02:00 on 5 November.\\
b. The time at which the CME is at $21.5\rsun$.\\
c. Radial velocity in km s$^{-1}$\\
\label{table_nariaki}
\end{table}

\section{Discussion and Conclusions}
\label{S9}

The solar events of 4 November 2015 present an ideal case for a multi-instrument and multi-event study with similar, closely spaced in time, events that are different in detail. They also provide a very good  opportunity to address some of the current discrepancies in the observations and theory of solar activity. All three events had associated M-class X-ray flares, EUV waves, CMEs, and CME-driven shocks. The first two events occurred near the western limb above AR 12445, which showed very rapid emergence (in just three to four days). They were homologous with similar characteristics but also had differences. Their EUV waves were mostly directed southward from the active region, with the brightest parts being low above the solar disk. Any shocks likely occurred close to the source (consistent with the weak Moreton wave observed), and along the dominant direction of propagation southward from the source region. The first outburst had a strong microwave flare, observed by the Nobeyama radioheliograph and polarimeters; its X-ray and microwave properties point to a very strongly localised and low source, consistent with the high starting frequency for a Type II burst. Event 2, however, had only a weak microwave burst but came from the same active region. 

Events 1\,--\,3 each produced a Type II burst, none of them identical or simple. Each can be interpreted in terms of multiple-lanes and/or split-bands and both fundamental and harmonic emission, often with very different intensities. Event~3 can be interpreted in terms of at least three Type II bursts or one Type II burst with at least six multiple lanes and split-bands. The Type II (and III) bursts observed should be excellent tests of theory, as well as the current understanding of their association with flare, CME, and SEP phenomena. 

Events 1 and 2 had unusually high-frequency Type II bursts, either high-frequency Type III bursts cutoff below $\approx 100$~MHz (Event 1) or no metric Type III bursts (Event 2), and minimal interplanetary Type III emission. These characteristics, especially the relative lack of interplanetary Type III activity, suggest that the associated shocks and flare sites were, despite both producing compact HXR sources seen by RHESSI, unable to produce accelerated particles that could either i) escape effectively into the high corona and interplanetary space, or ii) effectively produce radio emission in these regions, or iii) both. The absence of Type~III emission at meter and longer wavelengths has long been recognized as a typical property of strong flares in the western solar hemisphere that are not accompanied by SEP events \citep{Kle:al-10, Kle:al-11}. We emphasize nevertheless that electron beams can be present but radio-quiet in some frequency domains and radio-loud in others \citep{li_and_cairns_2012, li_and_cairns_2013}, as observed here for Event 3.

It is important to outline the connection between the dynamics of filament eruptions, EUV waves, CMEs, and Type II radio bursts for the homologous Events 1 and 2. Given the observational evidence, we provide one possible scenario. With the onset of the flares, relatively bright, impulsive EUV fronts were launched both on-disk and off-limb, reaching quite high speeds ($\gtrsim 900$~km~s$^{-1}$) early in the eruptions. The primarily south to southwest propagation direction for both was likely due to a ``funnel'' of weaker magnetic fields in that direction. This funnel may have focused the energy of the eruptions, allowing them to reach the high EUV intensity and speeds observed. The Type II bursts for both Events 1 and 2 began shortly after the onset of these bright, southward-directed EUV waves, indicating that the waves may have steepened into shocks quite early on and produced the radio emission. The Type II burst of Event 1 was much wider in frequency coverage than that of Event 2, indicating simultaneous emission from a range of coronal densities/heights, which agrees broadly with the EUV observations of the southern EUV front. The Type II burst continued for about ten minutes after the EUV front had left the AIA field of view. The EUV front of Event 2 had a smaller instantaneous height extent, was dimmer, and the associated Type II burst was narrower in frequency range and duration. We thus find strong temporal correlation, as well as qualitative correspondence between both the characteristics of the EUV and radio emission, and the scale of the two Types of emission, by comparing the two events.

Shortly after the onset of these seemingly impulsive EUV fronts, they were followed by much slower but persistent partial filament eruptions \,--\, initially in the southwest direction, eventually turning to the West. These drove the coronal waves in the west- and north-west directions, which eventually appeared as the bright CME fronts in coronagraphs. However, the slow filament eruptions only added to the fronts' energy for a short period of time, their peak speeds approaching the weak CME speeds, and decreasing afterwards. This may explain why the CMEs were weak, slow, and short-lived \,--\,  especially the CME of Event 2, which only had a failed filament eruption drive it. The slowness of the CMEs, despite the existence of fast wave-like disturbances in the low corona (the EUV waves), may be why the CMEs did not produce any Type II bursts in the high corona, as the Type II emission ended before the CMEs appeared in the LASCO C2 field of view. In the case of Event 2, the CME appeared in C2 a full half-hour after the Type II burst had ended.

Another important point is that the filament eruptions and EUV waves for Events 1 and 2 propagated primarily in the south to southwest directions, whereas the corresponding CMEs moved primarily west to northwest. This may have been due to the southwest-propagating off-limb portions of the EUV waves being decelerated by the overdense coronal plasma of the streamer, in which the flare site was embedded. At the same time, the radial portion of the compressive front kept propagating out through less dense material, additionally being driven by the partial filament eruptions. All of the above considerations lead us to conclude that the weak shock waves in Events 1 and 2 occurred early in the events, in the low corona, and were likely co-spatial to the observed EUV wavefronts. No Type II emission was observed after the EUV waves subsided, the quickly dissipating CMEs being the only interplanetary signatures of the events. 

The {\it in-situ} particle observations show a lack of high-energy protons for the first two events, while energetic electrons were observed only for Event 1. This is despite the nominally good magnetic connections for both the flare and CME sources (see Section 6 and 7, however). The CMEs from the first two events were directed mostly westwards near the Ecliptic from the west-limb active region, and exhibited a typical dome-like structure, even if the EUV waves were observed to propagate predominantly southward.

Event~3, in contrast to Events~1 and 2, was associated with lower frequency Type II emission, a very long-lasting Type IV burst between at least $10$~GHz and a few tens of MHz, and with coronal and interplanetary Type III bursts throughout the period 13:40\,--\,14:30~UT. Most but not all of these Type IIIs continue from above $10$~MHz to below $1$~MHz, with many of those that are not continuous appearing to be radio-quiet in the approximate domain $4 -- 30$~MHz, but radio-loud at higher and lower frequencies. The first Type IIIs  started during the HXR and microwave bursts and can be connected qualitatively by timing observations, the theoretical SEP transport model of \citet{AguedaEtAl08}, and significantly longer, non-Parker, field lines to the first set of SEP electrons observed by the {\it Wind}/3DP instrument. There are quantitative difficulties, however, with field line lengths of $1.5 -- 1.9$~AU needed but available models (Figure \ref{fig:BoLi_1}) only yielding $\approx 1.1 -  1.2$~AU. The timing and two-step nature of the SEP electron profile despite having relatively continuous Type IIIs below $\approx 4$~MHz, at least some of which are not seen in the ORF\'{E}ES and DAM spectra below about $30$~MHz, and more than two injections in the transport model's best-fit solution all require further work. Consideration of better magnetic field connectivity models, the ICME environment, cannibalisation of Type III electron streams \citep{li_etal_2002}, and some electron beams being radio-quiet in some frequency domains and radio-loud in others \citep{li_and_cairns_2012, li_and_cairns_2013} appear necessary. Event~1 should also be modelled using this approach and associated implications drawn. 

Some of Event~3's Type III bursts appear to be restricted to frequencies below the frequencies of simultaneous Type II bursts. One interpretation of this is that the shock wave producing the Type II bursts is also an accelerator of the electrons that produce Type~III bursts, a possible alternative to the usual scenario that Type III electron beams originate in magnetic reconnection regions. Alternatively, the Type III streams originate below the Type II shock but only become radio-loud after crossing the shock \citep{li_and_cairns_2012}. Another alternative is that the ``missing'' Type III radiation is generated outside of the CME and shock plasma but is prevented from reaching the observers at $1$~AU by propagation effects associated with the high-density regions behind the shock and CME.

The third event's CME was a half-halo in the southern hemisphere from the central disk active region. It showed much higher speeds in LASCO data than the first two. It drove an interplanetary shock; both the CME and shock reached Earth unexpectedly quickly (in $\approx$ two days), were not robustly or adequately predicted by ENLIL, and caused a space weather event with a geomagnetic storm. Interestingly, the SXR peak fluence method of \citet{CSM:al-17} predicted Event 3's arrival well and should be tested more.   

While SEP electrons were observed for Event~1, SEP protons were not and no SEPs were observed for Event~2. The relative lack of significant SEPs for these western events, despite the nominally good magnetic connectivity, the relatively bright EUV waves, and the intense, high frequency Type II bursts, may be due to the strong deceleration and narrow extents of the CME-driven shocks and so their inability to produce significant particle energization in the low corona. This is supported by the observed EUV disturbances and CMEs being fast and slow, respectively. The absence of strong Type III bursts for these events is an indication that the flare-accelerated particles either remained confined in the low corona or else were present but radio-quiet. Event~1's SEP electrons provide direct evidence for the radio-quiet interpretation \citep{li_and_cairns_2012, li_and_cairns_2013}. In addition, clearly, if there were significant flare- or shock-accelerated SEP protons for Event~1 or any SEPs at all for Event~2 then they did not find their way to open field lines connected to L1. This might be due to propagation effects or to particle injection onto non-connected field lines. Our magnetic-mapping and proton-transport analyses provide good evidence that the magnetic connectivity was not nominal for Events 1 and 2. Additionally, the CME eruptions were not radial for these events, complicating the study of connection points between the shock, flare site, and 1~AU. More detailed modelling of these events is required to differentiate between these interpretations. Even now, however, the analyses cast significant doubt on the typical relevance of simple connection analyses.

Event 3 actually produced significant levels of both electron and proton SEPs near $1$~AU, consistent with the observation of a rather fast and broad CME and with Type III bursts that argue for particle escape from the active region. The SEP event had very different electron and ion behaviors: a prompt, two-step profile for the electrons with rapid rises (near 14:10 and 14:40~UT, respectively) and obvious (by factors of $10 -- 100$) flux increases for $35 --  500$~KeV electrons, versus prompt (starting around 15:00~UT), slow rise (peak near 20:00~UT), and subtle (by $< 50\%$) flux increase for $10 -- 140$~MeV ions, with some evidence for 1~MeV ion increases and little below that energy. Focusing on the protons, since the electron SEPs and Type IIIs are discussed earlier, the proton acceleration and transport model of \citet{Marsh2015} leads to surprisingly reasonable agreement with the existence, onset time, single peak nature, and qualitative size of Event~3's proton SEP event. It also predicts weak, near background, fluxes of $10 -- 60$~MeV SEP protons for Events 1 and 2 that cannot be ruled out yet observationally.  The model does not include the ICME environment or non-Parker field lines but shows strong magnetic connectivity effects, since changing the solar wind speed (and associated Parker connectivity) to $500$~km~s$^{-1}$ from the observed value $\approx 700$~km~s$^{-1}$  leads to the prediction of clearly observable SEPs for Events~1 and 2 but none for Event~3, inconsistent with the observations. Future modelling work must address these effects. 

In summary, the three events that occurred on 4 November 2015 show both similarities and differences from standard events and each other, despite having very similar interplanetary conditions and only two flare sites and CME genesis regions. They are therefore targets for further in-depth observational studies, and for testing both existing and new theories and models, of flares, CMEs, the acceleration and transport of energetic particles, Type II, III, IV, microwave, and SA bursts, and related SEPs. Comparing the remote and {\it in-situ} observations of the three events, it remains possible that two traits of CME-related SEP-rich events are having i) sustained Type II emission to low enough frequencies and ii) a sustained high-speed shock, in order to ensure sufficient energization of the particles. It is also possible that once the SEPs have gained enough energy, they can scatter efficiently perpendicularly to the magnetic field and so perfect magnetic connectivity is not required for them to reach the $1$~AU observer. However, the results of this article show that magnetic connectivity is often not nominal ({\it e.g.} well described by a Parker spiral) and that both flare and CME sources of SEPs exist and may co-exist. While many aspects of Event 3's SEPs can be explained by the models presented, multiple aspects of the foregoing plasma, radio, X-ray, and energetic particle phenomena remain unexplained in detail at this time. More elaborate descriptions of the coronal shock dynamics and dynamic magnetic connectivity conditions are necessary for the study of both early-stage and later SEPs, particularly for widely separated observers. In addition, the results of this work reveal the complexity and interrelation of the chain of phenomena associated with solar eruptions. They demonstrate the need for strong integration of {\it in-situ} and remote magnetic, spectroscopic, particle, radio, and X-ray observations of active regions, flares, CMEs, radio and X-ray emissions, and SEPs with advanced theoretical models in order to gain deeper and more correct understanding of these phenomena.

%
\begin{acknowledgments}
The authors thank ISSI Bern for their hospitality to and support of the International Team on ``The Connection Between Coronal Shock Wave Dynamics and Early SEP Production'', from which the majority of this work resulted. A.~Veronig acknowledges Austrian Science Fund (FWF) grant P27292-N20. D.~Lario was supported by NASA grants NNX15AD03G and NNX16AF73G and the NASA Program NNH17ZDA001N-LW. N.~Nitta acknowledges support from NSF grant AGS-1259549. This work utilizes data obtained by the Global Oscillation Network Group (GONG) program, managed by the National Solar Observatory, which is operated by AURA, Inc. under a cooperative agreement with the National Science Foundation. It also uses data acquired by instruments operated by the Big Bear Solar Observatory, High Altitude Observatory, Learmonth Solar Observatory, Udaipur Solar Observatory, Instituto de Astrof\'{\i}sica de Canarias, and Cerro Tololo Interamerican Observatory. 
Data from the SOHO/ERNE instrument was provided by the Space Research Laboratory at the University of Turku, Finland. The authors thank all groups providing data.
\end{acknowledgments}


\bibliographystyle{spr-mp-sola}
\bibliography{overview_51_8Oct19}


\end{article}
\end{document}